\documentclass[twocolumn]{autart}    

\usepackage{graphicx}          

\usepackage{amsmath,amssymb,amsfonts, mathabx}

\usepackage[T1]{fontenc}
\usepackage{textcomp}
\usepackage{cite}
\usepackage{xcolor}
\usepackage{subcaption}
\usepackage{booktabs}
\usepackage{dsfont}
\usepackage{algorithm}
\usepackage[noend]{algpseudocode}
\usepackage{multirow}

\newcommand{\ig}[1]{\textcolor{blue}{[IG: #1]}}

\newcommand{\omegapath}{\omega}
\newcommand{\Omegapath}{\Omega}
\newcommand{\Omegapathfin}{\Omega^\text{fin}}
\newcommand{\x}{\boldsymbol{{x}}} 
\newcommand{\w}{\boldsymbol{{w}}}
\newcommand{\omegapathx}{\omegapath_{x}}
\newcommand{\omegapathxbold}{\boldsymbol{\omegapath}_{x}}
\newcommand{\Omegapathx}{\Omega_{x}}
\newcommand{\Omegapathxfin}{\Omega_x^\text{fin}}
\newcommand{\controller}{\kappa}
\newcommand{\strategy}{\sigma}
\newcommand{\Setofpolicies}{\Sigma}
\newcommand{\adversary}{\xi}
\newcommand{\Adversary}{\Xi}

\newcommand{\kernel}{\mathcal{T}}

\newcommand{\reals}{\mathbb{R}}
\newcommand{\naturals}{\mathbb{N}}
\newcommand{\uint}{\naturals_0}

\newcommand{\I}{\mathcal{I}}
\newcommand{\U}{\mathcal{U}}

\newcommand{\safe}{\text{safe}}
\newcommand{\unsafe}{\text{unsafe}}
\newcommand{\Xreach}{X_{\text{reach}}}
\newcommand{\Xavoid}{X_{\text{avoid}}}
\newcommand{\Xsafe}{X_\safe}

\newcommand{\Sreach}{S_{\text{reach}}}

\newcommand{\st}{\text{SMDP}}
\newcommand{\imdp}{\text{IMDP}}
\newcommand{\mi}{\text{MIMDP}}
\newcommand{\twolayer}{2\text{IMDP}}
\newcommand{\condprob}{\theta}
\newcommand{\condprobset}{\Theta}

\newtheorem{lemma}{Lemma}
\newtheorem{proposition}{Proposition}
\newtheorem{corollary}{Corollary}
\newtheorem{definition}{Definition}
\newtheorem{remark}{Remark}
\newtheorem{example}{Example}

\newtheorem{problem}{Problem}

\begin{document}

\begin{frontmatter}

\title{
Beyond Interval MDPs:\\Tight and Efficient Abstractions of Stochastic Systems\thanksref{footnoteinfo}} 

\thanks[footnoteinfo]{This paper was not presented at any IFAC 
meeting.}
\thanks[corr]{Corresponding author.}

\author[boulder]{Ibon Gracia\thanksref{corr}}\ead{ibon.gracia@colorado.edu},    
\author[boulder]{Morteza Lahijanian}\ead{morteza.lahijanian@colorado.edu}               
\address[boulder]{Ann \& H.J. Smead Department of Aerospace Engineering Sciences, University of Colorado Boulder, 80302 Boulder, CO}  
          

\begin{abstract}                          

This work addresses the general problem of control synthesis for continuous-space, discrete-time stochastic systems with probabilistic guarantees via finite abstractions. While established methods exist, they often trade off accuracy for tractability. We propose a unified abstraction framework that improves both the tightness of probabilistic guarantees and computational efficiency.
First, we introduce multi-interval MDPs (MI-MDPs), a generalization of interval-valued MDPs (IMDPs), which allows multiple, possibly overlapping clusters of successor states. 
This results in tighter abstractions but with increased computational complexity.
To mitigate this,
we further propose a generalized form of MDPs with set-valued transition probabilities (SMDPs), which model transitions as a fixed probability to a state cluster, followed by a non-deterministic choice within the cluster, as a sound abstraction.
We show that control synthesis for MI-MDPs reduces to robust dynamic programming via linear optimization, while SMDPs admit even more efficient synthesis algorithms that avoid linear programming altogether. Theoretically, we prove that, given the partitioning of the state and disturbance spaces, both MI-MDPs and SMDPs yield tighter probabilistic guarantees than IMDPs, and that SMDPs are tighter than MI-MDPs.
Extensive experiments across several benchmarks validate our theoretical results and demonstrate that SMDPs achieve favorable trade-offs among tightness, memory usage, and computation time.
 

\end{abstract}

\end{frontmatter}

\section{Introduction}
\label{sec:introduction}


Stochastic systems serve as fundamental models for uncertain dynamical control systems, where ensuring \emph{probabilistic guarantees} is crucial for \emph{safety-critical} applications. However, providing such guarantees remains a major challenge, especially for systems with nonlinear dynamics or non-Gaussian disturbances. A powerful approach to this problem is formal verification or control synthesis via \textit{finite abstraction}, wherein a continuous-space stochastic system is approximated by a finite-state Markov process that explicitly accounts for discretization errors.
While existing abstraction methods have been successful in specific settings, they are often limited to particular classes of dynamics and suffer from the state-explosion problem, compromising both scalability and accuracy. This work aims to develop a general abstraction framework for discrete-time stochastic systems that improves both \emph{accuracy} and \emph{computational tractability}, enabling more precise and efficient computation of guaranteed probabilistic bounds.

Several works have studied stochastic abstractions for control synthesis with formal guarantees. These works first obtain a partition of the continuous state-space of the original system and then assign a state of the abstraction to each region in the partition. Such abstractions are Markov models, like Markov Decision Processes (MDPs)~\cite{esmaeil2013adaptive} and interval-valued MDPs (IMDPs) \cite{lahijanian2015formal, cauchi2019efficiency, laurenti2020formal, badings2023probabilities}. 
In the case of MDPs, the abstraction error is typically computed and propagated separately from the model, and then combined with the verification results, resulting in conservative guarantees. In contrast, IMDPs incorporate the error directly into the abstraction, leading to more accurate verification outcomes.
While most works assume simple dynamics, such as those that are affine in the state and disturbance \cite{lahijanian2015formal,cauchi2019efficiency, laurenti2020formal, badings2022sampling}, recent research leverages reachability computations to allow nonlinearities in the state \cite{dutreix2022abstraction,adams2022formal, dutreix2018efficient, nazeri2025data}, and in both state and disturbances \cite{skovbekk2023formal,gracia2024data,gracia2024temporal}. 

More recently, several works aim to improve the accuracy or computational tractability of abstraction-based approaches. Such works leverage clustering of regions \cite{skovbekk2023formal,gracia2024temporal}, optimal partitioning of the state space \cite{skovbekk2023formal}, and more informed abstractions in the form of uncertain MDPs (UMDPs)~\cite{gracia2024temporal, mathiesen2024scalable,coppola2024data}. In particular, \cite{skovbekk2023formal} proposes an IMDP abstraction for general stochastic systems (nonlinear dynamics and non-Gaussian noise) via partitioning of the disturbance space and by bounding the probability of transitioning to unions (clusters) of states, and introduces a value iteration algorithm that accounts for this additional information. They show that using clusters yield tighter results.  
Work \cite{gracia2024temporal} extends this idea by considered a $2$-layer partition of the state space: a fine one and a coarse one, where the latter consists of non-overlapping clusters of the fine regions. 
Then, this information is encoded into a UMDP abstraction, specifically referred to as $2$-layer IMDP or $2$-interval MDP ($2$I-MDP), and designs a tailored synthesis algorithm that accounts for all constraints in the abstraction.  
The results show that $2$I-MDPs 
produce tighter bounds in the satisfaction probabilities than IMDPs.  However, such abstractions only admit non-overlapping clusters of states, limiting their expressive power.

Another class of UMDPs that reasons about transitions to sets of states are MDPs with set-valued transition probabilities (SMDPs)~\cite{trevizan2007planning}, recently explored in \cite{yu2025planning} for planning under temporal logic specifications. In an SMDP, the model transitions to some cluster (set) with a given probability, and then the successor state is adversarially chosen from the cluster. However, these models have not been used as abstractions of continuous-state stochastic systems.




We note that, besides abstraction-based approaches, other works deal with complex, i.e., nonlinear and stochastic dynamics via barrier certificates \cite{salamati2021data,Mazouz:NeurIPS:2022, mathiesen2024data} or stochastic simulation functions \cite{lavaei2022constructing}. However, these approaches are more conservative than abstraction-based methods~\cite{laurenti2023unifying} and often limited to bounded-horizon properties.  


In this work, we introduce two abstraction frameworks that are both tighter and more efficient than IMDPs and 2I-MDPs, as proposed in~\cite{skovbekk2023formal,gracia2024temporal}. First, we relax the assumptions of~\cite{gracia2024temporal} by allowing multiple, arbitrarily shaped clusters, leading to a more expressive abstraction class we call multi-interval MDPs (MI-MDPs). MI-MDPs offer improved tightness but incur additional computational burden. To better balance accuracy and tractability, we also propose a second abstraction that generalizes the SMDPs introduced in~\cite{trevizan2007planning}. Despite being more expressive, our SMDPs retain the same efficient control synthesis algorithm as in~\cite{trevizan2007planning}.
We further show that for any partition of the state and disturbance spaces, SMDPs consistently yield tighter results than the IMDPs of~\cite{skovbekk2023formal}. Moreover, we prove that even when incorporating additional transition probability information, as in $2$I-MDPs~\cite{gracia2024temporal}, or allowing overlapping clusters as in MI-MDPs, the resulting abstractions are no tighter than those obtained via our SMDPs.

In short, the contribution of this work is five-fold:
\begin{itemize}

    \item Introduction of MI-MDPs as a generalization of IMDPS and $2$I-MDPs, allowing multiple overlapping clusters to yield tighter abstractions at the cost of increased computational complexity.

    \item Generalization and application of SMDPs as sound abstractions for stochastic systems, while preserving efficient control synthesis.

    \item Proof of tightness dominance of SMDPs for any given partitioning, showing that they provide tighter probabilistic guarantees than IMDPs and are at least as tight (if not more) as $2$I-MDPs and MI-MDPs under any choice of the clusters.

    \item Theoretical characterization of control synthesis complexity, showing that robust dynamic programming reduces to linear programming for MI-MDPs, and that SMDPs support an even more efficient algorithm, as in~\cite{trevizan2007planning}.

    \item Comprehensive trade-off analysis between tightness, memory, and computation time, supported by both theoretical results and empirical evaluations across abstraction classes.

\end{itemize}

\section*{Basic Notation}

For clarity, we let $\uint$ denote the set of non-negative integers $\naturals\cup\{0\}$. Given a set $X$, we denote by $\mathds{1}_X$ the indicator function of $X$, i.e., $\mathds{1}_X(x) = 1$ if $x \in X$ and $0$ otherwise. We define the binary function $\boldsymbol{1}:\{\top, \bot\} \rightarrow\{0, 1\}$, which returns $1$ if its argument is true ($\top$) and $0$ otherwise. We also denote by $\mathcal{P}(X)$ the set of Borel probability distributions (measures) on $X$. Given a Borel set $A \subseteq X$ and a distribution $P \in \mathcal{P}(X)$, we let 
$P(A)$ denote the measure (probability) of the event $A$. For conciseness, we write the probability of the singleton event $\{a\}$ as $P(a) \equiv P(\{a\})$. We also write the Dirac measure on $x$ as $\delta_x$, such that $\delta_x(X) = 1$ if $x \in X$ and $0$ otherwise. We use bold symbols to indicate random variables, e.g., $\x \in \mathbb{R}$ is a real-valued random variable, whereas $x \in \mathbb{R}$ is a point (outcome) in the sample space $\reals$ of $\x$. 
\section{Problem Formulation}
\label{sec:problem_formulation}

We consider discrete-time stochastic systems of the form
\begin{align}
\label{eq:sys}
    \x_{t+1} = f(\x_t, u_t, \w_t),
\end{align}
where $\x_t \in \reals^n$ is the state of the system at time $t \in \uint$, 
$u_t \in U \subset \reals^m$ with $|U| < \infty$ is the control input\footnote{
While $U$ is finite, each of its elements may represent a continuous set of controllers (e.g., a partition of a continuous set), making this assumption non-restrictive.}, 
and $\w_t \in W \subseteq \reals^d$ is the disturbance (noise).
We assume $(\w_t)_{t \in \uint}$ is an \emph{i.i.d.} stochastic process where each $\w_t$ is a sample from a given probability distribution $P_W$. Finally, vector field $f:\reals^n \times U \times W \rightarrow \reals^n$ is a (possibly nonlinear) function of all its arguments, with $f(x,u,\cdot)$ being measurable for all $(x,u)$-pairs. 

Given time horizon $T \in \naturals_0$, states $x_0,\ldots,x_T \in \mathbb{R}^n$, and controls $u_0,\dots,u_{T-1}\in U$, we define a finite \textit{trajectory} of System~\eqref{eq:sys} as $\omegapathx =x_0 \xrightarrow{u_0}
\ldots \xrightarrow{u_{T-1}} x_T$ with length $|\omegapathx| = T + 1$. 
We let $\Omegapathxfin$ and $\Omegapathx$ be the sets of trajectories of finite and infinite lengths, respectively, 
and denote the state of $\omegapathx$ at time $t$ by $\omegapathx(t)$. 

We define a \emph{controller} of System~\eqref{eq:sys} as a function $\kappa: \Omegapathxfin \to U$ that maps each finite trajectory $\omegapathx$ to a control $\kappa(\omegapathx) \in U$. 
Given a state-control pair $(x_t,u_t)$ and a Borel set $B \subseteq \reals^n$, the (measurable) \emph{transition kernel} $\kernel:\mathcal{B}(\reals^n)\times \reals^n\times U \rightarrow [0, 1]$
of System~\eqref{eq:sys} determines the probability that $x_{t+1} \in B$, i.e.,
$\kernel(B\mid x,u) = \int_W \mathds{1}_{B}(f(x,u,w))P_W(dw)$, where 
$P_W(c)$ is the probability measure of $c \in \mathcal{B}(W)$. 
Given a controller $\controller$
and an initial state $x_0\in\reals^n$, the kernel $\kernel$ defines a unique probability measure $\text{Pr}_{x_0}^{\controller}$ over the trajectories of System~\eqref{eq:sys}~\cite{bertsekas1996stochastic}. 


We aim to compute a controller for System~\eqref{eq:sys} that satisfies a complex temporal requirement over regions in $\reals^n$ with high probability. These specifications, often expressed in temporal logic (e.g., LTL, LTLf), reduce to \emph{reach-avoid} properties over an extended state space via a finite abstraction. For simplicity of presentation, in this work we focus on these properties.
Given the sets $\Xreach, \Xavoid \subseteq \reals^n$ with $\Xreach \subseteq (\reals^n\setminus \Xavoid) =: \Xsafe$, we denote by $\varphi_x \equiv (\Xreach, \Xavoid)$ a \textit{reach-avoid} specification, which requires reaching $\Xreach$ while avoiding $\Xavoid$.
The probability that System~\eqref{eq:sys} satisfies $\varphi_x$ under controller $\controller$ from an initial state $x_0 \in \reals^n$ is defined as
\begin{multline}
    \label{eq: satisfaction_prob}
    \text{Pr}_{x_0}^{\controller}[\varphi_x] = \text{Pr}_{x_0}^{\controller}\big(\big\{\omegapathxbold \in \Omegapathx \mid \exists t \in \mathbb{N}_0 \text{ s.t. }   
     \omegapathxbold(t)\in\Xreach \\
     \land \; \forall \,t' \le t, \; \omegapathxbold(t')\notin \Xavoid \big\} \big),
\end{multline}
%


To obtain an abstraction for System~\eqref{eq:sys} for the purposes of controller synthesis, a $\varphi_x$-conservative partition is needed.
\begin{definition}[$\varphi_x$-Conservative Partition]
    A finite partition $S = \{s_1, \ldots, s_{|S|-1}, s_\text{avoid}\}$ of $\reals^n$ is called $\varphi_x$-conservative 
    if 
    (i) $\cup_{s\in S_\safe} s \subseteq  \Xsafe$, where $S_\safe := \{s_1, \dots, s_{|C|-1}\}$, 
    (ii) $s_\text{avoid} \supseteq \Xavoid$, and 
    (iii) there exists a maximal subset $S_\text{reach} \subseteq S_\safe$ s.t. $\cup_{s \in S_\text{reach}} s \subseteq \Xreach$.
\end{definition}


A general abstraction model is UMDP, which subsumes all the existing models, e.g., IMDPs \cite{skovbekk2023formal}.
%
\begin{definition}[UMDP]
    \label{def:UMDP}
    A UMDP is a tuple $\U := (S,A,\Gamma)$ in which $S$ and $A := U$ are respectively finite set of states and actions, and $\Gamma := \{\Gamma_{s,a} : s\in S, a \in A\}$, where $\Gamma_{s,a}$ is the set of transition probability distributions, or ambiguity set, of the pair $(s, a)$.
\end{definition}

%
%

\begin{definition}[Sound UMDP Abstraction]
    \label{def:sound_MDP abstraction}
    Given a $\varphi_x$-conservative partition $S$, a UMDP abstraction $\U = (S, A, \Gamma)$ in Def.~\ref{def:UMDP} is \emph{sound} if (i) for every $s\in S_\safe$, $x \in s$, $a\in A$, the distribution $\gamma_{x,a}$ given by $\gamma_{x,a}(s') := \kernel(s' \mid x, a)$ for all $s'\in S$ satisfies $\gamma_{x,a}\in \Gamma_{s,a}$, and (ii) $\Gamma_{s_\text{avoid}, a} = \{\delta_{s_\text{avoid}}\}$ for all $a\in A$
    .
\end{definition}

In this work, we aim to generalize the construction of a sound abstraction by solely using the reachable set computation of System~\eqref{eq:sys}, similar to \cite{skovbekk2023formal,gracia2024temporal}.
\begin{definition}[$\mathrm{Reach}$]
    The $1$-step forward reachable set of $s\subseteq \reals^n$, $a \in A$, and $c \subseteq W$ is defined as $\mathrm{Reach}(s,a,c) := \{f(x,a,w) : x\in s, w \in c\}$.
\end{definition}
There exist numerous approaches to obtain (overapproximations) of $\mathrm{Reach}$, \cite{althoff2021set}. Hence, we assume $\mathrm{Reach}$ operator or its overapproximation\footnote{Overapproximation of $\mathrm{Reach}$ is sufficient for soundness but may increase conservatism.}, also denoted by $\mathrm{Reach}$, is available.
We now have all the ingredients to formalize our abstraction for control synthesis problems.
%
%

\begin{problem}[Abstraction for Synthesis]
\label{prob:problem}
    Given System~\eqref{eq:sys}, its $\mathrm{Reach}$ operator, reach-avoid property $\varphi_x = (\Xreach, \Xavoid)$, and $\varphi_x$-conservative partition $S$,
    \begin{enumerate}
        \item[I.] using the $\mathrm{Reach}$ operator, construct a sound UMDP abstraction $\U$, and
        \label{prob:problem1}
        \item[II.] using $\U$, synthesize controller $\kappa$ and high probability functions $\underline p, \overline p:\reals^n \rightarrow [0, 1]$ such that
        $\rm{Pr}_{x_0}^{\kappa}[\varphi_x] \in [\underline p(x_0), \overline p(x_0)]$ for all $x_0\in\reals^n$.
    
        \label{prob:problem2}
    \end{enumerate}    
\end{problem}


Problem~\ref{prob:problem} is well-studied, and several abstraction methods, mostly into IMDPs, already exist \cite{skovbekk2023formal, cauchi2019efficiency, adams2022formal}. Our approach, however, differs in that it provably provides a higher lower bound $\underline{p}(x_0) $ for $ \Pr_{x_0}^{\kappa} $ and a tighter error bound $ \overline{p}(x_0) - \underline{p}(x_0) $ than existing methods for the same partition $S$, without requiring refinement. 

The key advantage lies in obtaining an ambiguity set $\Gamma$ that more precisely captures uncertainty in the dynamics of System~\eqref{eq:sys}. Specifically, we propose to abstract System~\eqref{eq:sys} into two novel UMDP classes, namely, set-valued MDPs (SMDPs) and multi-interval MDPs
(MI-MDPs). 
Unlike IMDPs, which only bound the probability of transitioning to individual regions $s' \in S$, these models reason about the probability of transitioning to more complex regions, such as clusters of states, which leads to more accurate results.

\section{Preliminaries: UMDP Semantics}
\label{sec:preliminaries}



For a given UMDP $\U$, we define a \textit{path} $\omegapath =s_0 \xrightarrow{a_0}
\ldots \xrightarrow{a_{T-1}} s_T$ to be a sequence of states such that for all $0 \leq t \leq T$, $s_t \in S$, and for all $0 \leq t \leq T-1$, $a_t \in A$ and there exists distribution $\gamma \in \Gamma_{s_t,a_t}$ with $\gamma(s_{t+1}) > 0$. 
We let $\Omegapathfin$ and $\Omegapath$ be the sets of all paths of finite and infinite length, respectively.
A \emph{strategy} of $\U$ is a function $\strategy: \Omegapathfin \rightarrow A$ that maps each finite path to the next action. We denote by $\Sigma$ the set of all strategies of $\U$.  When the value of $\sigma$ only depends on the current state, it is denoted a \emph{stationary} strategy. Given a finite path $\omegapath \in \Omegapathfin$ with last state $s_t$ and a strategy $\strategy \in \Sigma$, $\U$ transitions from $s_{t}$ under $a_t = \strategy(\omegapath)$ to $s_{t+1}$ according to some probability distribution in $\Gamma_{s_t,a_t}$, which is chosen by the adversary \cite{givan2000bounded}. Formally, an \emph{adversary} is a function $\adversary: S \times A \times \naturals_0 \rightarrow \mathcal{P}(S)$ that maps each state $s_t$, action $a_t$, and time step $t$ to a transition probability distribution $\gamma\in\Gamma_{s_t,a_t}$, according to which $s_{t+1}$ is distributed. We let $\Adversary$ denote the set of all adversaries. Given an initial state $s_0\in S$, a strategy $\strategy\in\Setofpolicies$, and an adversary $\xi\in\Xi$, $\U$ collapses to a Markov chain, with a unique probability measure over its paths. With a slight abuse of notation, we also denote this measure by $\text{Pr}_{s_0}^{\strategy,\xi}$.

\section{Tight Uncertain Abstraction}
\label{sec:abstraction}

In this section, we show how to abstract System~\eqref{eq:sys} into both an MI-MDP and an SMDP, given a $\varphi$-conservative partition $S$ and a partition $C$ of the disturbance set $W$. We highlight that most existing approaches \cite{skovbekk2023formal,chatterjee2023learner,badings2024learning} consider such a partition $C$. On the other hand, although approaches that estimate transition probabilities from samples of $P_W$ do not require this partition, most of them propose to cluster the samples, by proximity, into a set of regions \cite{badings2023probabilities, gracia2024temporal, gracia2024data}, which is very similar to defining a partition $C$ of $W$. We start with MI-MDPs in Section~\ref{sec:multi_imdp_abstraction}, and show how this class generalizes IMDPs~\cite{skovbekk2023formal} and $2$-layer IMDPs (2I-MDPs)~\cite{gracia2024temporal}. We then analyze the challenges that arise with such abstractions, which motivates the introduction of SMDP abstractions, discussed in Section~\ref{sec:MDP_ST_abstraction}. In particular, 
we formally prove that SMDPs are at least as accurate as MI-MDPs and empirically show that they often perform better.



\subsection{Multi-Interval MDP Abstraction}
\label{sec:multi_imdp_abstraction}

Here, we present our approach to constructing an MI-MDP abstraction of System~\eqref{eq:sys}.
Our method is based on the following lemma, which shows how to bound the transitions of System~\eqref{eq:sys}
using the $\mathrm{Reach}$ operator.
%
\begin{lemma}[\!\!{\cite[Theorem 1]{skovbekk2023formal}}]
    \label{lemma:tp_bounds}
    Consider a region $s\in S_\safe$, an action $a\in A$, the partition $C$ of the disturbance set $W$, and Borel set $r \subseteq \reals^n$. 
    Then, the transition kernel from each $x \in s$ to region $r$ under action $a$ satisfies
    $\kernel(r \mid x, a) \in [\underline P(s, a, r), \overline P(s, a, r)]$, where\footnote{We require the intersection of any (random) reachable set with a Borel set to be measurable. As shown in \cite{gracia2024temporal}, Lipschitz continuity of $f$ on $w$, uniformly over $x$, is sufficient (see \cite[Assump.~2.1]{gracia2024temporal})}
    \begin{subequations}
        \label{eq:tp_bounds}
        \begin{align}
            &\underline P(s, a, r) := \! \sum_{c \in C} \boldsymbol{1}(\mathrm{Reach}(s,a,c) \subseteq r)P_W(c),
            \label{eq:tp_bounds_lower}\\
            &\overline P(s, a,r) := \! \sum_{c \in C} \boldsymbol{1}(\mathrm{Reach}(s,a,c) \cap r \neq \emptyset)P_W(c).
             \label{eq:tp_bounds_upper}
        \end{align}
    \end{subequations}
\end{lemma}
%

Using these bounds, we define our MI-MDP abstraction.
\begin{definition}[MI-MDP Abstraction]
    \label{def:multi_imdp_abstraction}
    For each state-action pair $(s, a) \in S_\safe \times A$, let $\widetilde S_{s,a}\subseteq 2^{\reals^n}$ 
    be a set of (possibly overlapping) unions (clusters) of regions in $S$, i.e., each $\tilde s \in \widetilde S_{s,a}$ can be written as $\tilde s := \bigcup_{i = 1}^m s_i$, for some $s_1, \dots, s_m \in S$. We define the \emph{Multi-Interval MDP} (MI-IMDP) abstraction of System~\eqref{eq:sys} as a tuple $\U^\mi = (S,A,\Gamma^\mi)$, where 
    \begin{multline}
        \label{eq:Gamma_poly}
        \Gamma_{s,a}^\mi := \big\{ \gamma \in \mathcal{P}(S) : \: \forall \tilde s \in \widetilde S_{s,a},\\
        \underline P(s,a,\tilde s) \le \sum_{s'\in\{s''\in S : s'' \subseteq \tilde s\}} \gamma(s') \le \overline P(s,a,\tilde s)
         \big\}
    \end{multline}
for all $s\in S_\safe$ and $a\in A$, where $\underline P(s,a,\tilde s)$ and $\overline P(s,a,\tilde s)$ are defined in~\eqref{eq:tp_bounds}, and $\Gamma_{s_\text{avoid}, a}^\mi := \{\delta_{s_\text{avoid}}\}$ for all $a \in A$
%
\end{definition}
%


Intuitively, the MI-MDP abstraction is similar to IMDPs in that both represent uncertainty using bounds on transition probabilities between elements of $S$. The key difference is that while IMDPs consider bounds on transitions from each region to every other individual region in $S$, MI-MDPs instead define bounds on transitions to various \emph{clusters} of regions. In this way, MI-MDP generalizes both IMDPs and 2I-MDPs: when each cluster $\tilde s$ is equal to a single region $s \in S$, the MI-MDP reduces to an IMDP; when the clusters consist of  regions in $S$ as well as another set of non-overlapping unions of regions in $S$, 
it reduces to a 2I-MDP.

The additional constraints on feasible transition probability distributions in MI-MDPs generally yield a tighter and less conservative representation of the original system's dynamics compared to standard IMDPs. The number and size of the clusters in an MI-MDP can be user-defined. 

We illustrate all these abstractions in the example below.




%

\begin{figure}[]
    \centering
    \begin{subfigure}[t]{0.8\linewidth}
        \centering
        \includegraphics[width=0.8\linewidth]{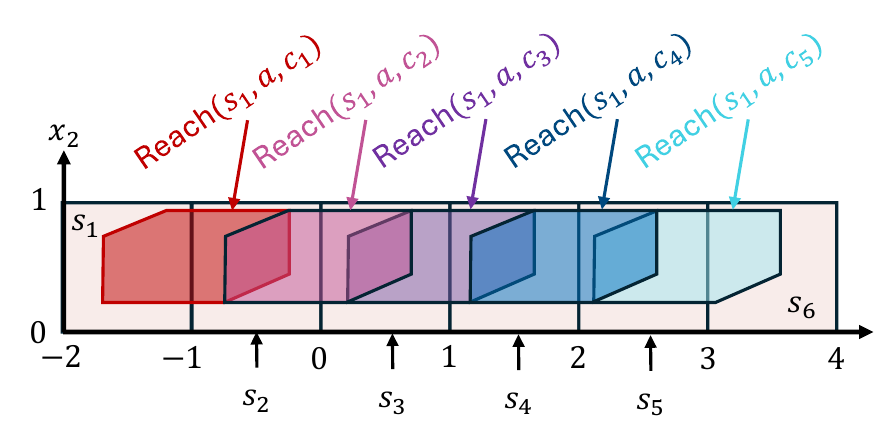}
        \caption{Discretization \& reachable sets
        }
        \label{fig:example_a}
    \end{subfigure}
    \hfill
    \begin{subfigure}[t]{0.49\linewidth}
        \centering
        \includegraphics[width=\linewidth]{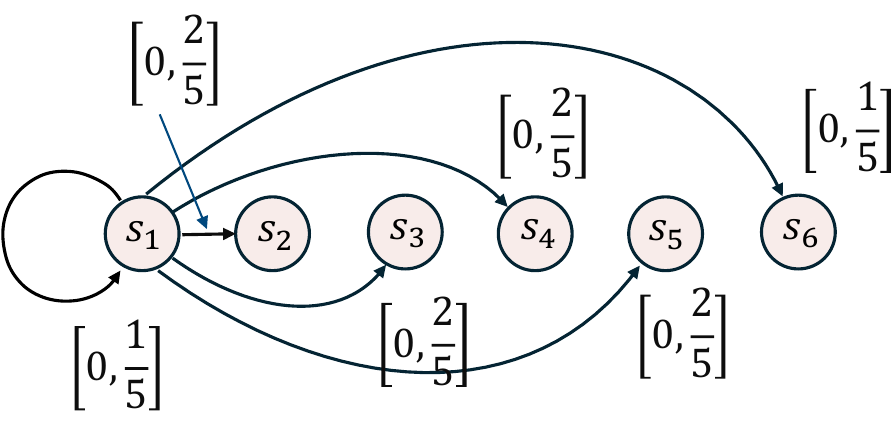}
        \caption{IMDP}
        \label{fig:example_b}
    \end{subfigure}  
    \begin{subfigure}[t]{0.49\linewidth}
        \centering
        \includegraphics[width=\linewidth]{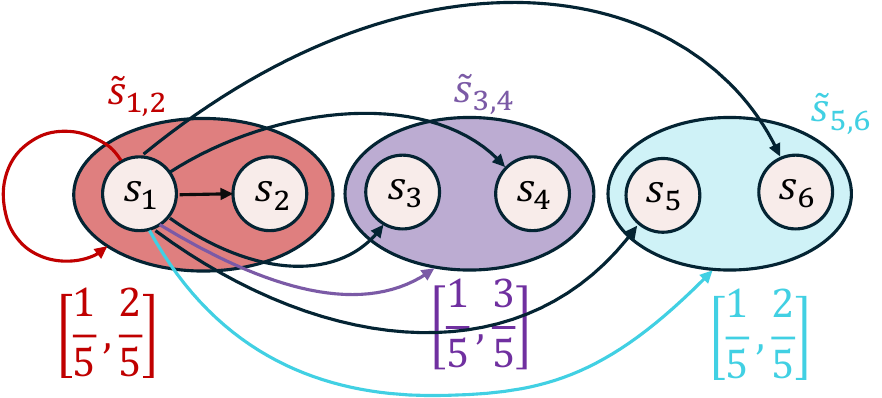}
        \caption{$2$I-MDP}
        \label{fig:example_c}
    \end{subfigure}
    \hfill
    \begin{subfigure}[t]{0.49\linewidth}
        \centering
        \includegraphics[width=\linewidth]{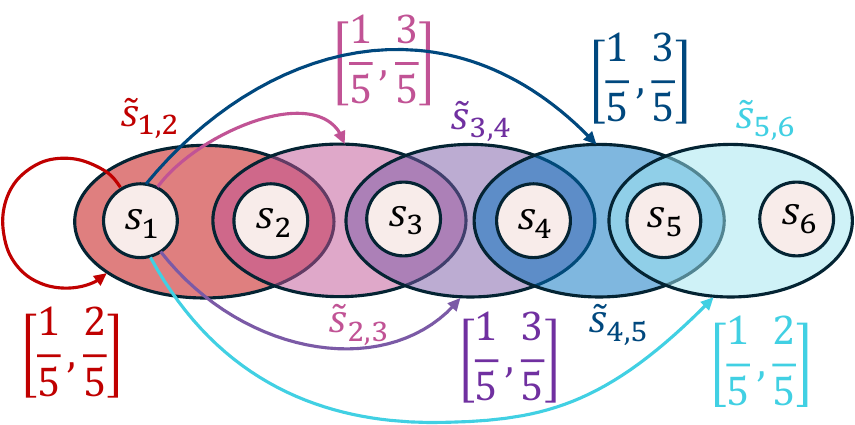}
        \caption{MI-MDP}
        \label{fig:example_d}
    \end{subfigure}
    \hfill    
    \begin{subfigure}[t]{0.49\linewidth}
        \centering
        \includegraphics[width=\linewidth]{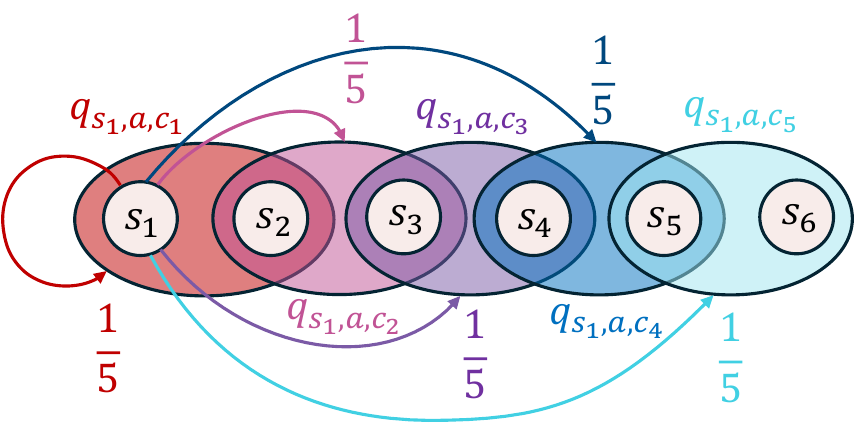}
        \caption{SMDP}
        \label{fig:example_e}
    \end{subfigure}\caption{\ref{fig:example_a}--\ref{fig:example_d} illustrate the setup and the different abstractions discussed in Examples~\ref{ex:conservativeness_imdps} and \ref{ex:example2}. For clarity, in \ref{fig:example_c} and \ref{fig:example_d}, we omit transition probability intervals that are shared with the IMDP in \ref{fig:example_b}. \ref{fig:example_d} shows the MI-MDP abstraction with informed clusters, whereas \ref{fig:example_d} shows the SMDP abstraction.
    }
    \label{fig:myfig}
\end{figure}

\begin{example}
\label{ex:conservativeness_imdps}
    Consider the \emph{linear time-invariant} system
    \begin{align*}
        \x_{t+1} = \begin{bmatrix}
            0.5 & 0.2\\
            0 & 0.5
        \end{bmatrix}\x_t + \begin{bmatrix}
            0.25\\
            0.7
        \end{bmatrix}u_t + \begin{bmatrix}
            0\\
            2.4
        \end{bmatrix}\w_t,
    \end{align*}
    with a single control $U = \{a\}$, $a = 1$. Let $P_W$ be the uniform distribution on $W = [-1, 1]$, which is partitioned uniformly into 5 regions $C = \{c_1, \ldots, c_5\}$, where $c_1 = [-1, -0.6]$, $c_2 = [-0.6, -0.2]$, $c_3 = [-0.2, 0.2]$, $c_4 = [0.2, 0.6]$ and $c_5 = [0.6, 1]$. The state partition $S=\{s_1,\ldots,s_6\}$, and reachable sets   $\mathrm{Reach}(s_1,a,c_i)$ are shown in Figure~\ref{fig:example_a}. 
    We obtain the IMDP abstraction $\U^\imdp = (S,A,\Gamma^\imdp)$ per Definition~\ref{def:multi_imdp_abstraction} and setting $A = U$ and $\widetilde S_{s_1,a} = S$.  The transitions of the IMDP from $s_1$ are shown in Figure~\ref{fig:example_b}. Note that 
    $\underline P(s_1, a, s') = 0$ for all states $s' \in S$ because $\mathrm{Reach}(s_1,a,c) \not \subseteq s'$ for each $c \in C$ as shown in Figure~\ref{fig:example_a}. Furthermore, $\overline P(s_1,a,s_1) =  \frac{1}{5}$ because only $1$ reachable set intersects $s_1$. Similarly, 
    the upper bounds for all the other transition probabilities are obtained.

    Now consider the transitions from $s_1$ in Figure~\ref{fig:example_b}.
    The distribution $\gamma^\imdp$ such that $\gamma^\imdp(s_1) = \frac{1}{5}$, $\gamma^\imdp(s_2) = \gamma^\imdp(s_3) = \frac{2}{5}$ and $\gamma^\imdp(s_4) = \gamma^\imdp(s_5) = \gamma^\imdp(s_6) = 0$ satisfies the transition probability bounds, and therefore is valid, i.e., $\gamma^\imdp \in \Gamma^\imdp_{s_1,a}$. However, assigning zero mass to the states $s_5$ and $s_6$ is a behavior that the original system cannot exhibit: for any $x_t \in s_1$, the probability that $\x_{t+1}$ lands in either $s_5$ or $s_6$ should be at least $
    \frac{1}{5}$ since the probability that $\w_t \in c_5$ is $P(c_5) = \frac{1}{5}$ and the reachable set $\mathrm{Reach}(s_1, a, c_5)$ of $c_1$ is fully contained in $s_5\cup s_6$. Therefore, the probability that the outcome of $\w_t$ generates a successor state $\x_{t+1}$ in $s_5$ or $s_6$ is no less than $\frac{1}{5}$.
    

    Because the IMDP abstraction includes such spurious behaviors, it often yields overly conservative results. Therefore, it is beneficial to consider an abstraction that encodes information about the probability of transitioning to unions (clusters) of regions, such as $\tilde s_{5,6} = s_5\cup s_6$, which would yield the constraint $\gamma(s_5) + \gamma(s_6) \ge \underline P(s_1,a, \tilde s_{5,6}) = \frac{1}{5}$.
    
    One option is to leverage a $2$-layer discretization \cite{gracia2024temporal} where the coarse layer contains the clusters $\tilde s_{1,2} = s_1 \cup s_2$, $\tilde s_{3,4} = s_3 \cup s_4$ and $\tilde s_{5,6} = s_5 \cup s_6$, yielding the $2$I-MDP $\U^{\twolayer}$ in Figure~\ref{fig:example_c} (only the additional transitions are shown). This $\U^{\twolayer}$ contains less spurious distributions than $\U^\imdp$, resulting in a tighter ambiguity set. For instance, $\gamma^{\imdp}$ is no longer a feasible distribution in $\U^{\twolayer}$ since $\gamma^{\imdp}(s_5) + \gamma^{\imdp}(s_6)=0$, which violates the constraint $\underline P(s_1,a,\tilde s_{5,6}) = \frac{1}{5}$.
    
    However, note that $\U^{\twolayer}$ is still not fully free from spurious distributions. For instance, let $\gamma^{\twolayer}$ be given by $\gamma^{\twolayer}(s_2) = \gamma^{\twolayer}(s_3) = \frac{2}{5}$, $\gamma^{\twolayer}(s_6) = \frac{1}{5}$, and $0$ otherwise. 
    This distribution satisfies all bounds in Figure~\ref{fig:example_c}, and therefore $\gamma^{\twolayer} \in \Gamma^{\twolayer}_{s_1,a}$ holds. However, $\gamma^{\twolayer}$ assigns zero probability to states $s_4$ and $s_5$, which the original system cannot: since $\mathrm{Reach}(s_1,a,c_4) \subset s_4\cup s_5$ and $P(c_4) = \frac{1}{5}$, the probability that $\x_{t+1}$ lands in either $s_4$ or $s_5$ is no less than $\frac{1}{5}$. 
    In fact, this is an inherent problem with 2I-MDP abstractions because it is typically unclear how to define a coarse discretization that is both non-overlapping and which reduces conservatism as much as possible. 
    
    On the other hand, one can let the clusters of the discretization be informed by the system's dynamics, and allow them to be overlapping. A possibility is to define each cluster as the union of the regions that intersect each reachable set, which we call \emph{informed} clustering. This leads to an MI-MDP $\U^\mi$ as in Definition~\ref{def:multi_imdp_abstraction}, with clusters 
    $\tilde s_{i,i+1} = s_i\cup s_{i+1}$ for all $i \in \{1,\ldots, 5\}$.
    This $\U^\mi$ is shown in Figure~\ref{fig:example_d}. Note that $\U^\mi$ also includes the IMDP transitions in Figure~\ref{fig:example_b} but are omitted in~Figure~\ref{fig:example_d} for visual clarity. 
    Note that $\gamma^{\twolayer}$ is no longer a feasible distribution in $\U^\mi$ since $\gamma^{\twolayer}(s_4)=\gamma^{\twolayer}(s_5)=0$, which violates the constraint $\underline P(s_1,a,\tilde s_{4,5}) = \frac{1}{5}$.

\end{example}

As illustrated in Example~\ref{ex:conservativeness_imdps}, it is beneficial to consider abstractions that account for the probability of transitioning to various clusters of regions. 
%
It however comes at the cost of increased computational complexity, both in computing transition kernel bounds 
as discuss in Section
~\ref{sec:memory}. Moreover, selecting the number and size of clusters introduces a trade-off between abstraction tightness and computational tractability, with no clear method on how to choose these parameters optimally. As we further demonstrate in Section~\ref{sec:synthesis}, synthesizing controllers for general MI-MDPs can be computationally demanding. 
To address these challenges, we now introduce an alternative abstraction model that avoids the need to bound the transition kernel via intervals.

\subsection{Set-valued MDP Abstraction}
\label{sec:MDP_ST_abstraction}

We introduce SMDPs as an alternative abstraction framework that reasons about transitions to sets of states, but which does so differently from MI-MDPs.  Instead of interval-valued transition probabilities, an SMDP specifies only one (single-valued) transition probability to each cluster for a given $(s,a)$ (see Figure~\ref{fig:example_e}). 
Once a cluster is reached, the distribution of the successor state $s'$ inside the cluster is chosen non-deterministically from the set of all conditional distributions on that cluster.

Moreover, SMDPs address the challenge of selecting appropriate clusters in MI-MDPs by automatically defining the clusters such that a single-valued transition probability is obtained.   
This is achieved simply by assigning probability $P_W(c)$ to the cluster of states $s' \in S$ that intersect with $\mathrm{Reach}(s,a,c)$ (e.g., see Figure~\ref{fig:example_e}).
To emphasize this distinction and improve clarity of presentation, we denote the clusters used for SMDPs by $q$, in contrast to $\tilde s$ notation used for MI-MDPs. With this intuition, we formalize our SMDP abstraction below.

\begin{definition}[SMDP Abstraction]
    \label{def:mdp_st_abstraction}
    For all $s\in S_\safe$, $a\in A$, $c\in C$, define cluster $q_{s,a,c} := \{s'\in S : s'\cap \mathrm{Reach}(s,a,c) \neq \emptyset \}$, and $Q_{s,a} := \{q_{s,a,c}: c\in C\}$. For each cluster $q_{s,a,c} \in Q_{s,a}$, let $\condprob(\cdot \mid q_{s,a,c}) \in \mathcal{P}(q_{s,a,c})$ denote a conditional probability distribution over the states in $q_{s,a,c}$ such that, for $s' \in q_{s,a,c}$, $\condprob(s'\mid q_{s,a,c})$ is the probability that the successor state of $(s,a)$ is $s'$ given that the transition to $q_{s,a,c}$ is realized.
    %
    Furthermore, let 
    $$\condprob_{s,a} := \big(\condprob(\cdot \mid q_{s,a,c}) \big)_{c \in C} \in \bigtimes_{c\in C} \mathcal{P}(q_{s,a,c}) =: \condprobset_{s,a}$$ 
    be an assignment of a conditional probability to each cluster in $Q_{s,a}$.
    %
    Finally,
    denote by $\gamma_{\condprob_{s,a}} \in \mathcal{P}(S)$ the distribution induced 
    by $\condprob_{s,a} \in \condprobset_{s,a}$ such that, for $s' \in S$, $\gamma_{\condprob_{s,a}}(s')$ is the (total) probability that the successor is $s'$, i.e.,
    \begin{align}
        \label{eq: SMDP exact successor prob}
        \gamma_{\condprob_{s,a}}(s') = \sum_{c \in \{c' \in C : s' \in q_{s,a,c'} \}} \condprob(s' \mid q_{s,a,c}) P_W(c).
    \end{align}
    %
    
    We define the SMDP abstraction of System~\eqref{eq:sys} as $\U^\st := (S,A,\Gamma^\st)$, where, for all $a \in A$,
    \begin{align}
    \label{eq:Gamma_mdp_st}
    \Gamma_{s,a}^\st := \{\gamma_{\condprob_{s,a}} : \condprob_{s,a} \in \condprobset_{s,a}\}
    \qquad \forall s\in S_\safe,
    \end{align}
    %
    and $\Gamma^\st_{s_\unsafe,a} := \{\delta_{s_\unsafe}\}$.
\end{definition}

SMDPs are similar to MI-IMDPs with informed clusters in that in both abstractions
the clusters are defined by taking into account the regions that intersect the reachable sets (e.g., see  Figures~\ref{fig:example_d} and \ref{fig:example_e}). 
The key difference is that MI-MDP abstractions bound the probability of transitioning from some $(s,a)$ to a cluster by counting how many reachable sets intersect or are subsets of the cluster, whereas SMDP abstractions make use of the fact that the reachable set $\mathrm{Reach}(s,a,c)$ has probability $P_W(c)$ of being realized, which implies that 
the probability of transitioning to the cluster $q_{s,a,c} \subseteq S$ is $P_W(c)$.
Then, to determine the probability of transitioning to $s' \in q_{s,a,c}$, the conditional probability distribution $\condprob(\cdot \mid q_{s,a,c})$ is needed.  This distribution however is uncertain in the same way that the exact transition probabilities in MI-MDPs are uncertain.
In fact, since the clusters $q_{s,a,c}$ for all noise partitions $c \in C$ can be overlapping, a conditional distribution $\condprob(\cdot \mid q_{s,a,c})$ for every $q_{s,a,c}$ that contains $s'$ is needed to determine the exact probability of transitioning to the successor state $s'$ per~\eqref{eq: SMDP exact successor prob}.

Hence, the state of an SMDP evolves as follows: from state $s$, the strategy chooses action $a$.  Next, the adversary chooses a feasible distribution $\gamma_{s,a} \in \Gamma_{s,a}^\st$ by picking a conditional distribution $\condprob_{s,a,c}$ per cluster $q_{s,a,c}$.  Then, the process transitions to $s' \in S$ with probability $\gamma_{s,a}(s')$ in~\eqref{eq: SMDP exact successor prob}.  


We note that the SMDP in Definition~\ref{def:mdp_st_abstraction} differs from the one in \cite{trevizan2007planning}.  In the latter, once the transition to a cluster $q'$ is realized, the successor state $s'$ is picked \emph{non-deterministically} from $q'$. In our SMDP, however, we allow probabilistic choices of $s'$ according to any (conditional) distribution in $\mathcal{P}(q')$.  
In fact, the incorporation of conditional distributions 
is key in establishing the soundness of SMDP abstractions (see Theorem~\ref{theorem:soundness} in Section~\ref{sec:analysis}). Nevertheless, in Theorem~\ref{thm:RVI_mdp_st}, we show that it suffices to consider only Dirac delta distributions for $\condprob(\cdot \mid q')$ (i.e., non-deterministic conditional choices of $s'$) to compute bounds on the probability of satisfying $\varphi$ in SMDPs.

%

We demonstrate our SMDP abstraction through the following example.
\begin{example}
\label{ex:example2}
    Consider again the setting of Example~\ref{ex:conservativeness_imdps}, and let $\U^\st = (S ,A, \Gamma^\st)$ be the corresponding SMDP abstraction. Figure~\ref{fig:example_e} shows the clusters $Q_{s_1, a} = \{q_{s_1,a,c_1}, \dots, q_{s_1,a,c_5}\}$, corresponding to transitions from $(s_1,a)$, where $q_{s_1,a,c_i} = \{s_{i}, s_{i+1}\}$ for all $i=\{1,\ldots,5\}$.
    Per Definition~\ref{def:mdp_st_abstraction}, there is a one-to-one correspondence between the clusters and the reachable sets, e.g., $q_{s_1,a,c_1}$ contains all and only the regions $s'\in S$ that intersect $\mathrm{Reach}(s_1,a, c_1)$. Furthermore, the probability of transitioning from $(s_1, a)$ to $q_{s_1,a,c_1}$ is given as $P_W(c_1) = \frac{1}{5}$, instead of as an interval, this being the case in MI-MDPs (see Figure~\ref{fig:example_d}).
    
    Within cluster $q_{s_1,a,c_1}$,
    the successor state is distributed according to some conditional distribution $\condprob(\cdot \mid q_{s_1,a,c_1})$. For instance, $\condprob(s_1 \mid q_{s,a,c_1}) = \condprob(s_2 \mid q_{s,a,c_1}) = \frac{1}{2}$ means that, if a transition to $q_{s,a,c_1}$ is realized, $s_1$ and $s_2$ are equally likely to be the successor state of $(s_1, a)$. Let the conditional distributions to the remaining clusters be $\condprob(\cdot \mid q_{s,a,c_2}) = \delta_{s_2}$, $\condprob(\cdot \mid q_{s,a,c_3}) = \delta_{s_3}$, $\condprob(\cdot \mid q_{s,a,c_4}) = \delta_{s_4}$ and $\condprob(\cdot \mid q_{s,a,c_5}) = \delta_{s_5}$. Then the total probability distribution $\gamma_{\condprob_{s_1,a}}$ of the successor state is uniquely defined as $\gamma_{\condprob_{s_1,a}}(s_1) = \frac{1}{10}$, $\gamma_{\condprob_{s_1,a}}(s_2) = \frac{3}{10}$, $\gamma_{\condprob_{s_1,a}}(s_i) = \frac{1}{5}$ for $i\in \{3,4,5\}$, 
    and $\gamma_{\condprob_{s_1,a}}(s_6) = 0$. 

    Note that the spurious distributions $\gamma^\imdp$ and $\gamma^{\twolayer}$ defined in Example~\ref{ex:conservativeness_imdps} are not allowed by $\U^\st$ (similar to MI-MDP), as they are not in $\Gamma_{s_1,a}^\st$: by~\eqref{eq: SMDP exact successor prob} and since $q_{s_1,a,c_5} = \{s_5, s_6\}$ we obtain that $\gamma^\imdp(s_5) + \gamma^\imdp(s_6)$ should be lower bounded by $\theta(s_5 \mid q_{s,a,c_5})P_W(c_5) + \theta(s_6 \mid q_{s,a,c_5})P_W(c_5) = \frac{1}{5}$, which is not the case. Similarly, since $q_{s_1,a,c_4} = \{s_4, s_5\}$, $\gamma^{\twolayer}(s_4) + \gamma^{\twolayer}(s_5)$ does not satisfy the lower bound $\theta(s_4 \mid q_{s,a,c_4})P_W(c_4) + \theta(s_5 \mid q_{s,a,c_4})P_W(c_4) = \frac{1}{5}$.

    Finally, we remark that, although the probability $P_W(c_i)$ of transitioning to a cluster $q_{s,a,c_i}$ is fixed, the uncertainty in the SMDP abstraction lies in the conditional probability distributions $\condprob(\cdot \mid q_{s,a,c_i})$, which
    can take any choice in $\mathcal{P}(q_{s,a,c_i})$. Such uncertainty gives rise to the set $\Gamma_{s_1,a}^\st$, and stems from the discretization of the state space and disturbance space. 
\end{example}

\section{Analysis of the Abstractions}
\label{sec:analysis}

In this section, we first show that both proposed abstraction classes MI-MDPs and SMDPs are sound abstractions of System~\eqref{eq:sys}, and that SMDPs represent the dynamics of the system at least as tightly as (if not more than) MI-MDPs for the same partitions $S$ and $C$, irrespectively of the choice of the clusters $\widetilde S_{s,a}$ of the MI-MDP. 
Then, we analyze the memory complexity of each abstraction. 

\subsection{Soundness and Tightness}
\label{sec:soundness_tightness}

We first prove soundness of SMDP abstractions.

\begin{thm}[Soundness of SMDP Abstraction]
\label{theorem:soundness}
    The SMDP $\U^\st$ obtained per Definition~\ref{def:mdp_st_abstraction} is a sound abstraction of System~\eqref{eq:sys}.
\end{thm}
\begin{pf*}{Proof.}
    Let $s\in S_\safe$, $a\in A$, 
    and pick $x\in s$. By the law of total probability, we obtain that, for all $s'\in S$, $\kernel(s' \mid x, a) = \sum_{c\in C} P_W(\{w \in c : f(x,a,w) \in s'\})$. Note that if $s' \notin q_{s,a,c}$, i.e., if $\mathrm{Reach}(x,a,c) \cap s' = \emptyset$, then $f(x,a,w) \notin s'$ for all $w \in c$, and thus $P_W(\{w \in c : f(x,a,w) \in s'\}) = 0$. Therefore, 
    we obtain $\sum_{c\in C} P_W(\{w \in c : f(x,a,w) \in s'\}) = \sum_{c\in \{c'\in C : s' \in q_{s,a,c}\}} P_W(\{w \in c : f(x,a,w) \in s'\})$. For all $s' \in S$ and $c \in C$, let $\condprob(s' \mid q_{s,a,c})$ denote the conditional probability of $f(x,a,w) \in s'$ given that $w \in c$, i.e., $\condprob(s' \mid q_{s,a,c}) := P_W(\{w \in c : f(x,a,w) \in s'\})/P_W(c)$. Note that $\condprob_{q_{s,a,c}}$ is supported on $q_{s,a,c}$, since 
    $\sum_{s'\in q_{s,a,c}} \condprob(s' \mid q_{s,a,c}) = \sum_{s' \in q_{s,a,c}}P(\{ w \in c : f(x,a,w) \in s' \})/P_W(c) = P(\cup_{s'\in q_{s,a,c}}\{ w \in c : f(x,a,w) \in s' \})/P_W(c) = P_W(\{ w \in c : f(x,a,w) \in \cup_{s'\in q_{s,a,c}} s' \})/P_W(c) = 1$, which holds due to $S$ being a partition, $f$ being deterministic and by definition of $q_{s,a,c}$. Therefore, $\kernel(s' \mid x, a) = \sum_{c\in \{c'\in C : s' \in q_{s,a,c}\}} \condprob(s' \mid q_{s,a,c}) P_W(c)$, implying that $\kernel(\cdot \mid x, a)\in \Gamma^\st_{s,a}$ $\forall x\in s$, which concludes the proof.
    %
    \qed
\end{pf*}

Next, we prove that our MI-MDP abstraction is also sound for System~\eqref{eq:sys}. We do this by using Theorem~\ref{theorem:soundness}.  That is, since SMDPs are sound, it suffices to prove that SMDPs are tighter abstractions than MI-MDPs, which is a key result of this work. 

\begin{thm}
\label{thm:multi_imdp_vs_mdp_st}
    Consider the fixed partitions $S$ and $C$ of the state and disturbance spaces. Let $\U^\st = (S,A, \Gamma^\st)$ and $\U^\mi = (S,A,\Gamma^\mi)$ 
    be respectively SMDP and MI-MDP abstractions of System~\eqref{eq:sys} per Definitions~\ref{def:mdp_st_abstraction} and~\ref{def:multi_imdp_abstraction}, respectively, where the sets $\widetilde S_{s,a}$ of clusters of $\U^\mi$ are arbitrary. Then, it holds that, $\Gamma_{s,a}^\st \subseteq \Gamma_{s,a}^\mi$ for all $s\in S$ and $a \in A$.

\end{thm}
\begin{pf*}{Proof.}
    Let $s\in S_\safe$ and $a \in A$. Given $\gamma^\st \in \Gamma_{s,a}^\st$, we prove that $\gamma^\st \in \Gamma_{s,a}^\mi$. Note that $\gamma^\st \in \Gamma_{s,a}^\st$ implies existence of conditional distributions $\condprob(\cdot \mid q_{s,a,c}) \in \mathcal{P}(q_{s,a,c})$ for all $c \in C$ such that $\gamma^\st(s') = \sum_{c \in C}\mathds{1}_{q_{s,a,c}}(s')\condprob(s' \mid q_{s,a,c})P_W(c)$ for all $s'\in S$. Let $s\in S_\safe$, $a\in A$ and $\tilde s\in \widetilde S_{s,a}$. We first prove that $\underline P(s,a,\tilde s) \le \sum_{s'\in\{s''\in S : s''\subseteq \tilde s\}} \gamma^\st(s')$. A sufficient condition, noting that $\gamma^\st(s') = \sum_{c \in C}\mathds{1}_{q_{s,a,c}}(s') \condprob(s' \mid q_{s,a,c}) P_W(c)$, is that $\boldsymbol{1}( \mathrm{Reach}(s,a,c) \subseteq \tilde s ) \le \sum_{s' \in\{s''\in S: s'' \subseteq \tilde s\}}\mathds{1}_{q_{s,a,c}}(s') \newline \condprob(s' \mid q_{s,a,c}) =  \sum_{s' \in\{s''\in q_{s,a,c}: s'' \subseteq \tilde s\}}\condprob(s' \mid q_{s,a,c})$ holds for all $c \in C$. Fix $c \in C$ and note that, if $\mathrm{Reach}(s,a,c) \subseteq \tilde s$, then $s'' \subseteq \tilde s$ for all $s'' \in q_{s,a,c}$, and therefore it holds that $\sum_{s' \in\{s''\in q_{s,a,c}: s'' \subseteq \tilde s\}}\condprob(s' \mid q_{s,a,c}) = \sum_{s' \in q_{s,a,c}}\condprob(s' \mid q_{s,a,c}) = 1 = \boldsymbol{1}( \mathrm{Reach}(s,a,c) \subseteq \tilde s )$. On the other hand, if $\mathrm{Reach}(s,a,c) \not \subseteq \tilde s$, then $\sum_{s'\in \{s''\in q_{s,a,c}: s'' \subseteq \tilde s\}} \condprob(s' \mid q_{s,a,c}) \ge 0 = \boldsymbol{1}( \mathrm{Reach}(s,a,c) \subseteq \tilde s )$. Since this holds for all $c\in C$, it follows that $\underline P(s,a,q) \le \sum_{s'\in q} \gamma^\st(s')$. Next, we follow the previous logic to prove that $\overline P(s,a,\tilde s) \ge \sum_{s' \in\{s''\in S: s'' \subseteq \tilde s\}} \gamma^\st(s')$. Fix $c\in C$ and note that, if $\mathrm{Reach}(s,a,c) \cap \tilde s \neq \emptyset$, then $\sum_{s' \in\{s''\in q_{s,a,c}: s'' \subseteq \tilde s\}} \condprob(s' \mid q_{s,a,c}) \le 1 = \boldsymbol{1}( \mathrm{Reach}(s,a,c) \cap \tilde s \neq \emptyset )$. On the other hand, if $\mathrm{Reach}(s,a,c) \cap \tilde s = \emptyset$, then $\sum_{s' \in\{s''\in q_{s,a,c}: s'' \subseteq \tilde s\}}\condprob(s' \mid q_{s,a,c}) = \sum_{s'\in q\cap q_{s,a,c}} \condprob(s' \mid q_{s,a,c}) = 0 = \boldsymbol{1}( \mathrm{Reach}(s,a,c) \cap \tilde s \neq \emptyset )$. Since this holds for all $c\in C$, we obtain $\overline P(s,a,q) \ge \sum_{s'\in q} \gamma^\st(s')$. Since the previous bounds on $\sum_{s' \in\{s''\in S: s'' \subseteq \tilde s\}}\gamma^\st(s')$ hold for all $\tilde s\in \widetilde S_{s,a}$, we obtain that $\gamma^\st \in \Gamma_{s,a}^\mi$, which concludes the proof.
    \qed
\end{pf*}

\begin{corollary}[Soundness of MI-MDP abstraction]
\label{cor:soundness_mimdp}
    The MI-MDP $\U^\mi$ obtained per Definition~\ref{def:multi_imdp_abstraction} is a sound abstraction of System~\eqref{eq:sys}.
\end{corollary}

\begin{remark}
\label{rem:tightness}
    The main reason behind the improved tightness of the ambiguity set in SMDPs compared to MI-MDPs is that
    the SMDP abstraction leverages the fact that the probability of $\w_t \in c$ is also the probability that $\mathrm{Reach}(s,a,c)$ is realized. In consequence, the semantics of the SMDP enforce that a set-valued transition to $q_{s,a,c}$ also has probability $P_W(c)$.     
    On the other hand, the MI-MDP abstraction does not leverage this information directly.  Instead, MI-MDP uses the reachable sets to bound the transition probabilities to each region (state), which leaves room for more spurious distributions. 
    Hence, even an MI-MDP abstraction constructed using the same (equivalent) clusters as in the SMDP abstraction may include spurious distributions that are not present in the SMDP model. 
    While this is not clear in Example~\ref{ex:example2}, with a slight modification, Example~\ref{ex:example3} below clearly illustrates this point.
    Therefore, in general $\Gamma_{s,a}^\st \subsetneq \Gamma_{s,a}^\mi$.
\end{remark}

\begin{figure}[]
    \centering
    \begin{subfigure}[t]{0.49\linewidth}
        \centering
        \includegraphics[width=\linewidth]{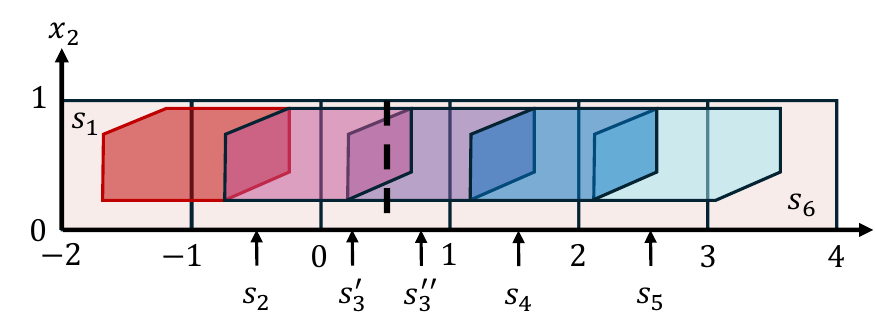}
        \caption{Discretization and reachable sets
        }
        \label{fig:example3_a}
    \end{subfigure}
    \hfill
    \begin{subfigure}[t]{0.49\linewidth}
        \centering
        \includegraphics[width=\linewidth]{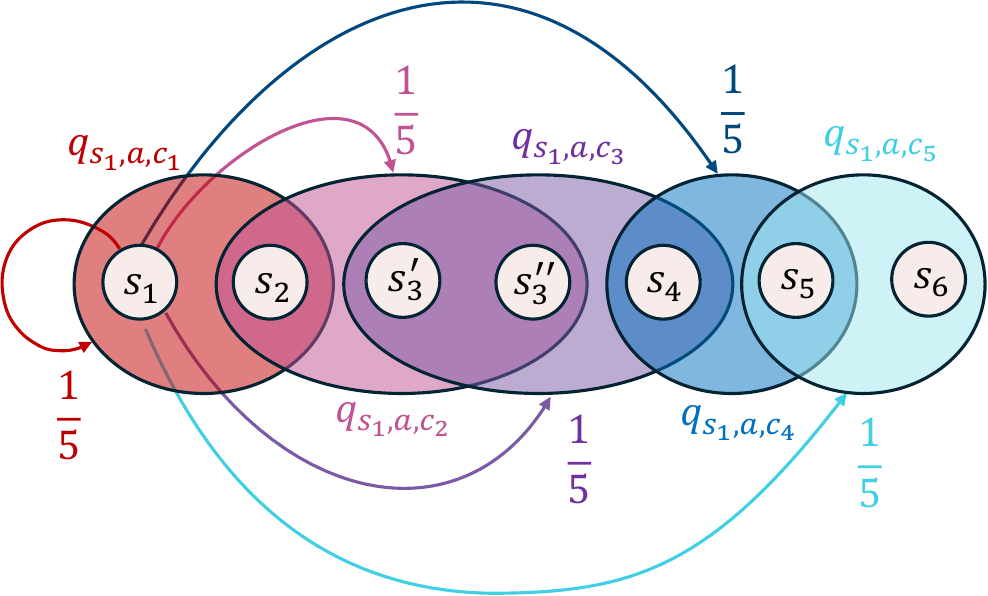}
        \caption{SMDP}
        \label{fig:example3_b}
    \end{subfigure}
    
    \begin{subfigure}[t]{\linewidth}
        \centering
        \includegraphics[width=0.49\linewidth]{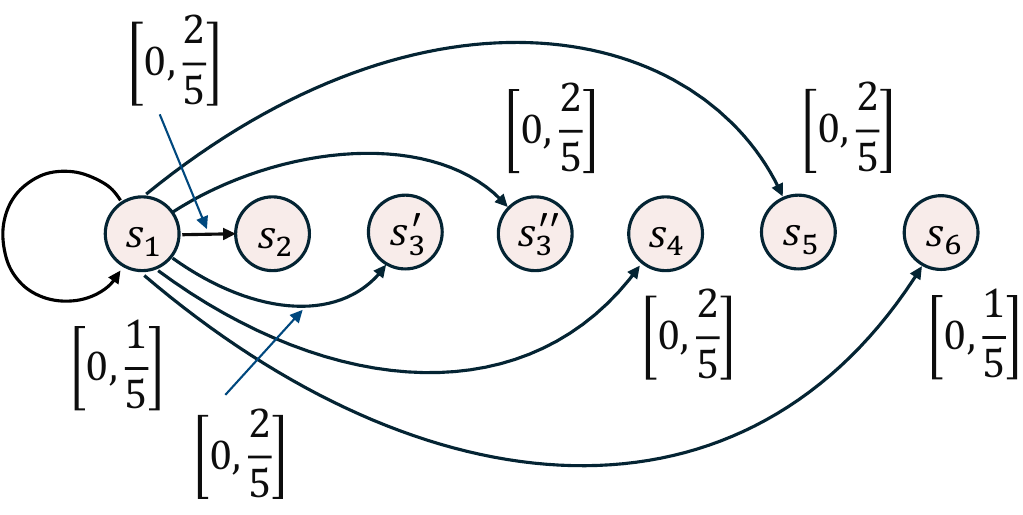}
        \includegraphics[width=0.49\linewidth]{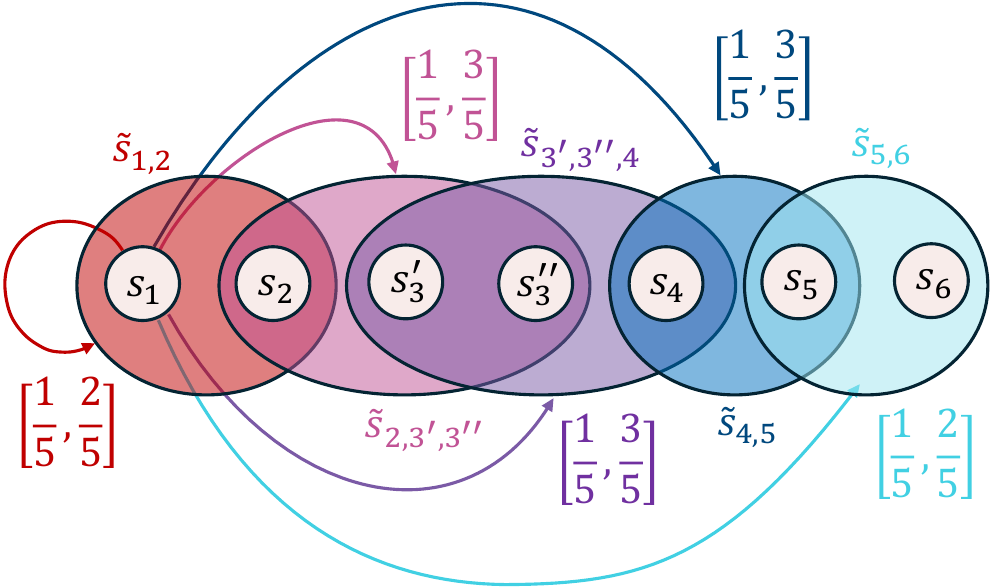}
        
        \caption{MI-MDP}
        \label{fig:example3_c}
    \end{subfigure}
    \hfill
    \caption{ (a) Setup of Example~\ref{ex:example3}. The SMDP abstraction is in (b), and MI-MDP abstraction is shown in (c) using two figures for clarity: (left) bounds for all states $s \in S$, and (right) bounds for all the (informed) clusters $\tilde s$.
    }
    \label{fig:myfig2}
\end{figure}

\begin{example}
    \label{ex:example3}
    Consider the same setup of Examples~\ref{ex:conservativeness_imdps} and \ref{ex:example2}, but let the state-space partition be $S = \{s_1, s_2, s_3', s_3'', s_4, s_5, s_6\}$, where the region $s_3$ is split into $s_3'$ and $s_3''$ as shown in Figure~\ref{fig:example3_a}. Figure~\ref{fig:example3_b} shows the SMDP abstraction per Definition~\ref{def:mdp_st_abstraction}, and Figure~\ref{fig:example3_c} shows the MI-MDP abstraction per Definition~\ref{def:multi_imdp_abstraction}, where the set of clusters $\widetilde S_{s,a}$ contains both informed clusters $\{\tilde s_{1,2}, \tilde s_{2,3,4}, \tilde s_{3,4,5}, \tilde s_{5,6}, \tilde s_{6,7}\}$ and the regions in $S$. 
    
    It is easy to observe that the distribution $\gamma^\mi$, with $\gamma^\mi(s_1) = \gamma^\mi(s_3') = \gamma^\mi(s_5) = \frac{1}{5}$, $\gamma^\mi(s_3'') = \frac{2}{5}$ and $\gamma^\mi(s_2) = \gamma^\mi(s_4) = \gamma^\mi(s_6) = 0$, satisfies all bounds in the transition probabilities, and therefore $\gamma^\mi \in \Gamma_{s_1,a}^\mi$. However, $\gamma^\mi$ does not belong to $\Gamma_{s_1,a}^\st$.  
    That is because, in the SMDP abstraction,
    it is impossible to obtain $\gamma^\mi(s_3') = \frac{1}{5}$ and $\gamma^\mi(s_3'') = \frac{2}{5}$,
    regardless the choice of the conditional distributions $\condprob(\cdot \mid q_{s_1,a,c_2})$ and $\condprob(\cdot \mid q_{s_1,a,c_3})$ of the SMDP as explained below. Observe that, for $\gamma^\mi(s_3'') = \frac{2}{5}$, it requires that $\condprob(\cdot \mid q_{s_1,a,c_2}) = \condprob(\cdot \mid q_{s_1,a,c_3}) = \delta_{s_3''}$. 
    However, this implies that $\condprob(s_3' \mid q_{s_1,a,c_2}) = \condprob(s_3' \mid q_{s_1,a,c_3}) = 0$, and therefore $\gamma^\mi(s_3'')$ can only be $0$. Since this is a contradiction, $\gamma^\mi \notin \Gamma_{s_1,a}^\st$, making it a spurious distribution of the MI-MDP abstraction.
    
\end{example}

\subsection{Computational Complexity of Abstraction}
\label{sec:memory}

Tight abstractions using  MI-MDPs and SMDPs, especially when compared to IMDPs, come at the cost of increased memory usage and computational effort.
This is because it is necessary to track the states associated with each cluster. When clusters overlap, the cost further increases, as some states belong to multiple clusters, hence raising the overall computational complexity.

The following propositions give the worst-case memory complexities of MI-MDPs and SMDPs as a function of the number of clusters and their sizes.
\begin{proposition}[SMDP Abst. Space Complexity]
    Let $\U^\st = (S,A,\Gamma^\st)$ be an SMDP  abstraction with the set of clusters $Q = \bigcup_{s \in S, a\in A} Q_{s,a}$,  
    and partition $C$ of the disturbance set $W$. 
    Denote by $N_q = \max_{q\in Q} |q|$ the size of the largest cluster in $Q$. Then, $\U^\st$ has a worst-case memory complexity of $\mathcal{O}(N_q|S|\, |A|\, |C|)$.
\end{proposition}

\begin{proposition}[MI-MDP Abst. Space Complexity]
    
    Let $\U^\mi = (S,A,\Gamma^\mi)$ be an MI-MDP abstraction with the set of clusters $\widetilde S = \bigcup_{s\in S, a\in A} \widetilde S_{s,a}$. 
    Let $N_{\tilde{s}} = \max_{\tilde{s} \in \widetilde{S}} \sum_{s \in S} \mathbf{1}(s \subseteq \tilde{s} )$ 
    be the maximum number of regions $s\in S$ that form a cluster $\tilde s$ of $\U^\mi$, 
    and denote by $N_{\widetilde{S}} = \max_{s\in S, a\in A}|\widetilde{S}_{s,a}|$ the size of the largest set of clusters.
    Then, $\U^\mi$ has a worst-case memory complexity of 
    $\mathcal{O}(N_{\tilde{s}}|S| \, |A| N_{\widetilde{S}})$.
\end{proposition}

Note that, given the partition $C$ of $W$, the complexity of an SMDP abstraction is fixed, as the number and size of the clusters is given in Definition~\ref{def:mdp_st_abstraction}. This is not the case in MI-IMDPs, where the number and size of the clusters is user defined. 


For example, let $\mathrm{Post}(s,a) := \{s'\in S: \exists \gamma \in \Gamma_{s,a}^\mi\:\text{s.t.}\:\gamma(s') > 0\}$ denote the set of possible successor states of an $(s,a)$-pair and $N_{\mathrm{Post}} = \max\{|\mathrm{Post}(s,a)| : s\in S, a\in A\}$. When for each $(s,a)$-pair, $\widetilde S_{s,a} = \mathrm{Post(s,a)}$, as is the case in IMDP abstractions, it holds that $N_{\tilde s} = 1$ and $N_{\overline S} = N_{\mathrm{Post}}$, and we recover the space complexity of IMDPs: $\mathcal{O}(|S||A|N_{\mathrm{Post}})$.

On the other hand, if we let the clusters of the MI-MDP be informed by the reachable sets and we do not consider transitions to individual regions, i.e., $\widetilde S_{s,a} := \{\cup_{s'\in q} s': q \in Q_{s,a}\}$, for all $s\in S$ and $a\in A$, where $Q_{s,a}$ is given in Definition~\ref{def:mdp_st_abstraction}, we recover the space complexity of SMDPs.
%
We highlight that, even though  for every $(s,a)$-pair, it holds that $\bigcup_{c\in C} q_{s,a,c} = \mathrm{Post}(s,a)$, generally $N_q |C| > N_{\mathrm{Post}}$ due to some clusters overlapping, and hence the memory complexity of SMDPs and MI-MDPs with informed clusters is typically higher than that of IMDPs.
Note that, in higher dimensions, this difference in required memory might be higher, as the overlap might be larger. 

Nevertheless, we also highlight that, unlike MI-MDPs, $2$I-MDPs, and IMDPs, which require storing transition probability bounds for all clusters, states and actions, the only probabilities that need to be stored in the case of SMDPs are the values $P_W(c)$ for all $c \in C$.  
This can result in significantly lower memory usage for SMDPs, particularly in low-dimensional state spaces, when $|S|$ is high and when $|C|$ is small. 
In Section~\ref{sec:case_studies} we empirically compare the memory requirements of each abstraction class as a function of the granularity of the partitions $S$ and $C$, and the dimension $n$ of the state space, validating this discussion.


Finally, note that an MI-MDP with the same space complexity as an SMDP does not necessarily yield an abstraction of comparable tightness. Achieving a tighter MI-MDP often requires going beyond reachable-set-informed clusters, which in turn increases the abstraction's space complexity. As discussed in Example~\ref{ex:example3}, one approach is to explicitly include the regions $s' \in \mathrm{Post}(s,a)$ in the cluster sets $\widetilde{S}_{s,a}$.
However, we remark that by Theorem~\ref{thm:multi_imdp_vs_mdp_st}, no matter the number or size of the clusters of the MI-MDP abstraction: the SMDP model will always be as tight (if not more), than the MI-MDP.


\section{Control Synthesis}
\label{sec:synthesis}


In this section, we present a method for synthesizing controllers that maximize the reach-avoid probability while being robust against all uncertainties embedded in the abstraction. Specifically, (i) we describe how to obtain strategies for general UMDPs as in Definition~\ref{def:UMDP},
which include MI-MDPs and SMDPs,
using Robust Dynamic Programming (RDP),
(ii) we prove that, specifically for these models, RDP reduces to solving linear programs (LPs)
at each iteration, and (iii) we show that SMDPs admit a very efficient tailored algorithm to solve these LPs.

\subsection{Strategy Synthesis via Robust Dynamic Programming}

Given UMDP $\U$, a reach-avoid specification $\varphi = (\Sreach, \{s_\text{avoid}\})$ with $\Sreach,\{s_\text{avoid}\}\subseteq S$, strategy $\sigma\in \Sigma$, and adversary $\xi\in\Xi$, we denote the reach-avoid probability from state $s \in S$ by $\text{Pr}_{s}^{\strategy,\xi}[\varphi]$, which is defined analogously to \eqref{eq: satisfaction_prob}.

\begin{proposition}[RDP {\cite[Theorem 6.2]{gracia2025efficient}}]
\label{prop:robust_value_iteration}
Given a UMDP $\U = (S,A,\Gamma)$, a reach-avoid specification $\varphi = (\Sreach, \{s_\text{avoid}\})$, and $s \in S$, 
define the optimal robust reach-avoid probability as 
$\underline p(s) :=\sup_{\sigma \in\Sigma}\inf_{\xi\in\Xi} \text{Pr}_{s}^{\strategy,\xi}[\varphi]$. Consider also the recursion
\begin{align}
    \label{eq:robust_value_iteration_lower_bound}
    \underline p^{k+1}(s) =\max\limits_{a\in A}\min\limits_{\gamma\in\Gamma_{s,a}} \sum\limits_{s'\in 
             S}\gamma(s')\underline p^{k}(s')
\end{align}
for all $s \in S\setminus \Sreach$, otherwise $\underline p^k(s) = 1$, where $k\in\uint$, with initial condition $\underline p^{0} = \mathds{1}_{\Sreach}(\cdot)$. Then, $\underline p^k$ converges to $\underline p$.
\end{proposition}

The major challenge in computing $\underline p$ is solving the inner minimization problems (over the sets $\Gamma_{s,a}$) in \eqref{eq:robust_value_iteration_lower_bound}. In Section~\ref{sec:inner_problems}, we show that these minimizations are linear programs (LPs). Based on the RDP in \eqref{eq:robust_value_iteration_lower_bound}, work \cite[Theorem 6.4]{gracia2025efficient} introduces a polynomial algorithm to obtain an optimal robust strategy , namely, which satisfies $\sigma^*(s) 
\in \arg\max_{a \in A}\min_{\gamma \in \Gamma_{s,a}} \sum_{s'\in S}\gamma(s')\underline p(s')$ for all $s\in S$, and which is also stationary
. Then using $\sigma^*$, we obtain the optimistic probabilities $\overline p(s) := \sup_{\xi\in\Xi} \text{Pr}_{s}^{\strategy^*,\xi}[\varphi]$ by iterating on the recursion in \cite[Equation 6.5]{gracia2025efficient}, which is similar to the one in \eqref{eq:robust_value_iteration_lower_bound} but where the $\min$ over $\Gamma_{s,a}$ is replaced by a $\max$, and the actions are determined by $\sigma^*$. Finally, we translate $\sigma^*$ to the controller $\kappa$ of System~\eqref{eq:sys} as $\kappa(x) := \sigma^*(s)$, with $ s \ni x$, for all $x\in \reals^n$. 

The following result provides the guarantees that System~\eqref{eq:sys} in closed loop with $\controller$ satisfies $\varphi_x$, thus solving Problem~\ref{prob:problem}.
\begin{thm}[Correctness of the Controller]
\label{thm:strategy_synthesis}
    Let $\underline p$ be obtained via the RDP recursion \eqref{eq:robust_value_iteration_lower_bound}, $\sigma^*$ be as per \cite[Theorem 6.4]{gracia2025efficient}, $\overline p$ be as in \cite[Equation 6.5]{gracia2025efficient} and $\kappa$ be obtained by refining $\sigma^*$ to System~\eqref{eq:sys}. 
    Then, for all $x_0\in \reals^n$, it holds that $\text{Pr}_{x_0}^{\controller}[\varphi_x] \in [\underline p(s), \overline p(s)]$ with $s\in S$ such that $x_0 \in s$.
\end{thm}

Representing uncertainty more tightly, SMDPs yield tighter results than MI-MDPs, which we now formalize.
\begin{thm}
\label{thm:tightness_guarantees}
    Let $\U^\mi$ and $\U^\st$ be respectively MI-MDP and SMDP abstractions of System~\eqref{eq:sys} obtained for the same discretizations $(S, C)$. Denote by $[\underline p^\mi, \overline p^\mi]$ and $[\underline p^\st, \overline p^\st]$ the satisfaction guarantees obtained for $\U^\mi$ and $\U^\st$ respectively. Then, for all $s\in S$, it holds that $[\underline p^\st(s), \overline p^\st(s)] \subseteq [\underline p^\mi(s), \overline p^\mi(s)]$.
\end{thm}
\begin{pf*}{Proof.}
    Start by assuming that, at iteration $k$ of RDP, the function $\underline p^{k+1}$ in the case of $\U^\st$ is pointwisely greater or equal than that of $\U^\mi$. Since $\Gamma^\st \subseteq \Gamma^\mi$ as established by Theorem~\ref{thm:multi_imdp_vs_mdp_st}, the solution $\underline p^{k+1}$ of \eqref{eq:robust_value_iteration_lower_bound} obtained on SMDPs is pointwisely dominates the one obtained on MI-MDPs. Since both sequences stat from the same initial condition, an induction argument shows that this dominance holds for every $k \geq 0$, i.e., the sequence $(\underline p^k(s))_{k \in \uint}$ obtained for an SMDP dominates, that of an MI-MDP, leading to a  higher $\underline p(s)$ for all $s\in S$. The same reasoning shows that $\overline p^\st(s) \le \overline p^\mi(s)$ for all $s\in S$, concluding the proof.
    \qed
\end{pf*}

\subsection{Inner Optimization Problems in \eqref{eq:robust_value_iteration_lower_bound}}
\label{sec:inner_problems}

Consider the inner minimization problems in \eqref{eq:robust_value_iteration_lower_bound}. Given the polytopic shape of the sets $\Gamma^\mi_{s,a}$ of an MI-MDP $\U^\mi$ in \eqref{eq:Gamma_poly}, it is easy to conclude that the inner minimization is a linear program. This means that synthesizing a controller for an MI-MDP via RDP boils down to solving LPs using standard solvers like GUROBI, which have complexity $\mathcal{O}(N_{\mathrm{Post}}^3)$. In consequence, the overall complexity of a single iteration of RDP is $\mathcal{O}(|S||A|N_{\mathrm{Post}}^3)$. Note that in the case that $\U^\mi$ is an IMDP or a $2$I-MDP, instead of LP, tailored algorithms are introduced in~\cite{givan2000bounded} and \cite{gracia2024temporal} that reduce this complexity to $\mathcal{O}(|S||A|N_{\mathrm{Post}}\log{(N_{\mathrm{Post}})})$. 

Below, we show that SMDP abstractions also admit tailored algorithms that eliminate the need to solve LPs, thereby significantly reducing computational complexity.

\begin{thm}[Inner Minimization for SMDPs]
\label{thm:RVI_mdp_st}
Consider an SMDP $\U^\st = (S,A, \Gamma^\st)$, and let $k\in \uint$, $s\in S_\safe$ and $a\in A$. Then, the inner problem in \eqref{eq:robust_value_iteration_lower_bound} is equivalent to: 
\begin{align}
\label{eq:mdp_ust_cost}
\min\limits_{\gamma\in\Gamma_{s,a}^\st} \sum\limits_{s'\in 
             S}\gamma(s')\underline p^{k}(s') = \sum_{c \in C}P_W(c) \min_{s' \in q_{s,a,c}}\underline p^{k}(s').
\end{align}
%
\end{thm}
\begin{pf*}{Proof.}
    Let $s\in S_\safe$ and $a\in A$. By the structure of $\Gamma^\st_{s,a}$ in \eqref{eq:Gamma_mdp_st}, the inner problem in \eqref{eq:robust_value_iteration_lower_bound} is equivalent to
    \begin{align*}
    \min_{\{\condprob(\cdot \mid q_{s,a,c}) \in \mathcal{P}(q_{s,a,c})\}_{c\in C}} \sum_{c \in C} \sum_{s' \in  q_{s,a,c}} \condprob(s' \mid q_{s,a,c}) P_W(c)  \underline p^{k}(s')\\
    \sum_{c \in C} P_W(c) \min_{\condprob(\cdot \mid q_{s,a,c}) \in \mathcal{P}(q_{s,a,c})}\sum_{s' \in  q_{s,a,c}} \condprob(s' \mid q_{s,a,c}) \underline p^{k}(s').
\end{align*}
Note that each $\min$ problem over $\condprob(\cdot \mid q_{s,a,c})$ is a linear program, since the objective is linear in $\condprob(\cdot \mid q_{s,a,c})$ and $\mathcal{P}(q_{s,a,c})$ is a polytope. As such, the optimal value is attained when each $\condprob(\cdot \mid q_{s,a,c})$ is at a vertex of $\mathcal{P}(q_{s,a,c})$, thus assigning probability $1$ to a single state in $q_{s,a,c}$, namely, the one with lowest $\underline p^k(s')$, and zero to all other states, which leads to \eqref{eq:mdp_ust_cost}. 
    \qed
\end{pf*}

The intuition behind Theorem~\ref{thm:RVI_mdp_st} is that, in order to minimize Expression~\eqref{eq:robust_value_iteration_lower_bound}, the adversary picks a $\gamma^* \in \Gamma^\st_{s,a}$ or, equivalently, the conditional distributions $\condprob(\cdot \mid q_1)^*,\dots, \condprob(\cdot \mid q_{|C|})^*$, in such a way that each $\condprob(\cdot \mid q_i)^*$ assigns probability $1$ to a single state $s' \in q_i$, namely, the one with the lowest $\underline p^k(s')$. Therefore, restricting the adversary to choosing $s' \in q_i$ deterministically is enough, which is the case of the set-valued MDPs in \cite{trevizan2007planning}. Consequently, we can perform RDP using Algorithm~\ref{alg:algo} \cite{trevizan2007planning}, which only requires performing finite searches.

\begin{algorithm}[t]\small
\caption{Inner minimization for SMDPs \cite{trevizan2007planning}}
\label{alg:algo}
\begin{algorithmic}
    \Require $\U^\st, \underline p^k, P_W$
    \Ensure $\underline p^{k+1}$
    \For{$s \in S$}
    \For{$a \in A$}
    \For{$c \in C$}
    \State $\underline p^k(q_{s,a,c}) \gets \min\{\underline p^k(s') : s' \in q_{s,a,c}\}$
    \EndFor
    \EndFor
    \State $\underline p^{k+1}(s) \gets \max \{\sum_{c\in C}P_W(c)\underline p^k(q_{s,a,c}) : a\in A\}$
    \EndFor
\end{algorithmic}
\end{algorithm}

\begin{proposition}
\label{prop:complexity_synthesis_smdp}
    Let $C$ be a partition of $W$ and $\U^\st = (S,A,\Gamma^\st)$ be the corresponding SMDP abstraction. 
    Then, the computational complexity of every iteration of RDP on $\U^\st$ is $\mathcal{O}(|S||A|N_q|C|)$.
\end{proposition}
\begin{pf*}{Proof.}
    By Theorem~\ref{thm:RVI_mdp_st}, each iteration of RDP requires finding, for each $s\in S, a\in A$ and $c\in C$, the minimum $\underline p^k(s')$ over the states $s'\in q_{s,a,c}$ via finite search, which has complexity $\mathcal{O}(N_q)$. The statement follows from this fact.
    \qed
\end{pf*}


From Proposition~\ref{prop:complexity_synthesis_smdp}, it follows that the ratio between computational complexity of control synthesis on SMDPs and that of IMDPs (and $2$I-MDPs) is $\mathcal{O}(N_q|C|/(N_{\mathrm{Post}}\log{(N_{\mathrm{Post}})}))$.
As discussed in Section~\ref{sec:memory}, the product $N_q|C|$ is often larger than $N_{\mathrm{Post}}$, which makes it difficult to provide a formal statement on which abstraction has lower complexity when it comes to control synthesis. Therefore, we just provide a qualitative analysis below. 

When the discretization $C$ is coarse and discretization $S$ is fine, or when the dimension $n$ of the state space is small, then $N_q |C|$ 
is not much larger than $N_{\mathrm{Post}}$. 
Hence, control synthesis in SMDPs becomes $\mathcal{O}(\log{(N_{\mathrm{Post}})})$ 
times cheaper than in IMDPs and $2$I-MDPs. As we show in Section~\ref{sec:case_studies}, under such conditions,
control synthesis is
up to one order of magnitude faster than in IMDPs and $2$I-MDPS. However, as $C$ becomes finer, $S$ becomes coarser, and $n$ increases,
control synthesis on SMDPs becomes increasingly more expensive than in IMDPs and $2$I-MDPs.

\begin{table*}[]
    \centering
    \caption{Benchmark results for all the case studies. 
    The evaluation metrics are: average error (tightness) $e_\text{avg}$ in Equation~\eqref{eq: average error}, abstraction time $T_\text{abs}$, synthesis time $T_\text{syn}$, memory usage to store the abstraction, and correctness through theoretical probability bounds $\underline{p}_\text{avg}$ and $\overline{p}_\text{avg}$ defined in Equation~\eqref{eq: average lower bound} and Monte Carlo simulation satisfaction probability $P^\kappa_\text{avg}[\varphi_x]$. 
    We set a timeout (TO) of $360$ minutes for $T_\mathrm{syn}$. 
    Underlined values denote changed parameters; bold indicates the best results.
    }
    \label{tab:experiment_results}
    \scalebox{0.94}{
    \begin{tabular}{c l c l c c c c c c c c c c}
        \toprule
        \multirow{2}{*}{\#} & \multirow{2}{*}{System} & \multirow{2}{*}{$n$} & \multirow{2}{*}{Abstraction} & \multirow{2}{*}{$|S|$} & \multirow{2}{*}{$|A|$} & \multirow{2}{*}{$|C|$} & \multirow{2}{*}{$e_\text{avg}$} & $T_\text{abs}$ & $T_\text{syn}$ & Memory & 
        \multicolumn{3}{c}{From Initial Set $S_{0}$}
        \\
        \cline{12-14}
         & &  &  &  &  &  &  & (min) & (min) & (GB) & $\underline p_\text{avg}$ & $\overline p_\text{avg}$ & $P^\kappa_\text{avg}[\varphi_x]$ \\
        \midrule
        $1$ & $2$D Unicycle & $2$ & IMDP & $\underline{901}$ & $8$ & $145$ & $0.156$ & $0.235$ & $\boldsymbol{0.620}$ & $0.017$ & $0.645$ & $1.000$ & $0.996$ \\
        $2$ &  &  & $2$I-MDP &  &  &  & $0.129$ & $0.359$ & $2.619$ & $0.018$ & $0.715$ & $1.000$ & $0.995$ \\
        $3$ &  &  & MI-MDP &  &  &  & $0.099$ & $1.872$ & $3.709$ & $0.045$ & $0.907$ & $1.000$ & $0.998$ \\
        $4$ &  &  & SMDP &  &  &  & $\boldsymbol{0.083}$ & $\boldsymbol{0.087}$ & $0.695$ & $\boldsymbol{0.014}$ & $\boldsymbol{0.923}$ & $1.000$ & $\boldsymbol{0.999}$ \\
        \cline{4-14}
        $5$ &  &  & IMDP & $\underline{2026}$ & $8$  & $145$  & $0.064$ & $0.597$ & $2.160$ & $0.076$ & $0.956$ & $1.000$ & $0.999$ \\
        $6$ &  &  & $2$I-MDP &  &  &  & $0.055$ & $0.853$ & $8.680$ & $0.082$ & $0.951$ & $1.000$ & $0.999$ \\
        $7$ &  &  & MI-MDP &  &  &  & $0.039$ & $4.086$ & $11.060$ & $0.148$ & $0.965$ & $1.000$ & $0.999$ \\
        $8$ &  &  & SMDP &  &  &  & $\boldsymbol{0.033}$ & $\boldsymbol{0.203}$ & $\boldsymbol{1.580}$ & $\boldsymbol{0.043}$ & $\boldsymbol{0.971}$ & $1.000$ & $0.999$ \\
        \cline{4-14}
        $9$ &  &  & IMDP & $\underline{3601}$ & $8$  & $145$  & $0.064$ & $1.280$ & $7.854$ & $0.236$ & $0.942$ & $1.000$ & $0.999$ \\
        $10$ &  &  & $2$I-MDP &  &  &  & $0.044$ & $1.707$ & $24.511$ & $0.255$ & $0.955$ & $1.000$ & $1.000$ \\
        $11$ &  &  & MI-MDP &  &  &  & $0.029$ & $7.258$ & $22.850$ & $0.387$ & $0.975$ & $1.000$ & $1.000$ \\
        $12$ &  &  & SMDP &  &  &  & $\boldsymbol{0.027}$ & $\boldsymbol{0.372}$ & $\boldsymbol{2.659}$ & $\boldsymbol{0.101}$ & $\boldsymbol{0.978}$ & $1.000$ & $1.000$ \\
        \cline{4-14}
        $13$ &  &  & IMDP & $\underline{5626}$ & $8$ & $145$ & $0.048$ & $2.395$ & $16.948$ & $0.577$ & $0.957$ & $1.000$ & $0.999$ \\
        $14$ &  &  & $2$I-MDP &  &  &  & $0.035$ & $2.960$ & $58.710$ & $0.626$ & $0.969$ & $1.000$ & $0.999$ \\
        $15$ &  &  & MI-MDP &  &  &  & $0.023$ & $11.504$ & $56.672$ & $0.856$ & $0.981$ & $1.000$ & $\boldsymbol{1.000}$ \\
        $16$ &  &  & SMDP &  &  &  & $\boldsymbol{0.020}$ & $\boldsymbol{0.603}$ & $\boldsymbol{4.455}$ & $\boldsymbol{0.205}$ & $\boldsymbol{0.984}$ & $1.000$ & $0.999$ \\
        \midrule
        $17$ & $3$D Unicycle & $3$ & IMDP & $27001$ & $10$ & $\underline{37}$ & $0.885$ & $2.800$ & $\boldsymbol{4.671}$ & $0.625$ & $0.048$ & $0.976$ & $0.866$ \\
        $18$ &  &  & $2$I-MDP &  &  &  & $0.884$ & $4.710$ & $15.005$ & $0.633$ & $0.050$ & $0.976$ & $0.868$ \\
        $19$ &  &  & MI-MDP &  &  &  & $0.740$ & $11.645$ & $347.826$ & $1.234$ & $0.218$ & $0.977$ & $0.910$ \\
        $20$ &  &  & SMDP &  &  &  & $\boldsymbol{0.558}$ & $\boldsymbol{1.325}$ & $10.563$ & $\boldsymbol{0.486}$ & $\boldsymbol{0.411}$ & $\boldsymbol{0.969}$ & $\boldsymbol{0.913}$ \\
        \cline{4-14}
        $21$ &  &  & IMDP &  $27001$ & $10$  & $\underline{65}$ & $0.710$ & $4.650$ & $\boldsymbol{4.769}$ & $\boldsymbol{0.586}$ & $0.243$ & $0.975$ & $0.874$ \\
        $22$ &  &  & $2$I-MDP &  &  &  & $0.704$ & $7.625$ & $18.060$ & $0.594$ & $0.247$ & $0.975$ & $0.877$ \\
        $23$ &  &  & MI-MDP &  &  &  & $-$ & $26.646$ & TO & $1.517$ & $-$ & $-$ & $-$ \\
        $24$ &  &  & SMDP &  &  &  & $\boldsymbol{0.367}$ & $\boldsymbol{2.322}$ & $14.228$ & $0.707$ & $\boldsymbol{0.605}$ & $\boldsymbol{0.968}$ & $\boldsymbol{0.910}$ \\
        \cline{4-14}
        $25$ &  &  & IMDP & $27001$ & $10$ & $\underline{101}$ & $0.497$ & $6.571$ & $\boldsymbol{6.765}$ & $\boldsymbol{0.554}$ & $0.479$ & $0.972$ & $0.880$ \\
        $26$ &  &  & $2$I-MDP &  &  &  & $0.490$ & $12.303$ & $22.283$ & $0.562$ & $0.485$ & $0.972$ & $0.876$ \\
        $27$ &  &  & MI-MDP &  &  &  & $-$ & $55.822$ & TO & $1.869$ & $-$ & $-$ & $-$ \\
        $28$ &  &  & SMDP &  &  &  & $\boldsymbol{0.260}$ & $\boldsymbol{3.450}$ & $24.414$ & $0.962$ & $\boldsymbol{0.707}$ & $\boldsymbol{0.963}$ & $\boldsymbol{0.909}$ \\
        \midrule
        $29$ & Temperature & $\underline{2}$ & IMDP & $\underline{145}$ & $1$ & $\underline{17}$ & $0.031$ & $\boldsymbol{0.0003}$ & $\boldsymbol{0.0009}$ & $6\times 10^{-5}$ & $0.974$ & $1.000$ & $1.000$ \\
         $30$ &  &  & $2$I-MDP &  &  &  & $0.030$ & $0.001$ & $0.002$ & $8\times 10^{-5}$ & $0.975$ & $1.000$ & $1.000$ \\
        $31$ &  &  & MI-MDP &  &  &  & $0.016$ & $0.002$ & $0.091$ & $10^{-4}$ & $0.986$ & $1.000$ & $1.000$ \\
        $32$ &  &  & SMDP &  &  &  & $\boldsymbol{0.013}$ & $\boldsymbol{0.0003}$ & $0.0021$ & $\boldsymbol{3\times 10^{-5}}$ & $\boldsymbol{0.989}$ & $1.000$ & $1.000$ \\
        \cline{3-14}
        $33$ &  & $\underline{3}$ & IMDP & $\underline{1729}$ & 1 & $\underline{65}$ & $0.066$ & $\boldsymbol{0.010}$ & $\boldsymbol{0.027}$ & $\boldsymbol{0.004}$ & $0.928$ & $1.000$ & $1.000$ \\
        $34$ &  &  & $2$I-MDP &  &  &  & $0.065$ & $0.018$ & $0.053$ & $\boldsymbol{0.004}$ & $0.940$ & $1.000$ & $1.000$ \\
        $35$ &  &  & MI-MDP &  &  &  & $0.039$ & $0.170$ & $1.958$ & $0.010$ & $0.966$ & $1.000$ & $1.000$ \\
        $36$ &  &  & SMDP &  &  &  & $\boldsymbol{0.015}$ & $\boldsymbol{0.010}$ & $0.099$ & $\boldsymbol{0.004}$ & $\boldsymbol{0.983}$ & $1.000$ & $1.000$ \\
        \cline{3-14}
        $37$ &  & $\underline{4}$ & IMDP & $\underline{20737}$ & 1 & $\underline{257}$ & $0.150$ & $0.740$ & $\boldsymbol{1.117}$ & $\boldsymbol{0.200}$ & $0.857$ & $1.000$ & $1.000$ \\
        $38$ &  &  & $2$I-MDP &  &  &  & $0.150$ & $1.245$ & $2.043$ & $0.205$ & $0.851$ & $1.000$ & $1.000$ \\
        $39$ &  &  & MI-MDP &  &  &  & $0.102$ & $39.257$ & $110.211$ & $0.828$ & $0.894$ & $1.000$ & $1.000$ \\
        $40$ &  &  & SMDP &  &  &  & $\boldsymbol{0.021}$ & $\boldsymbol{0.584}$ & $6.822$ & $0.544$ & $\boldsymbol{0.975}$ & $1.000$ & $1.000$ \\
        \bottomrule
    \end{tabular}
    }
\end{table*}

\section{Case Studies}
\label{sec:case_studies}

In this section we empirically evaluate the effectiveness of the proposed approaches to obtain UMDP abstractions and to synthesize controllers that yield tight satisfaction guarantees. We consider three case studies: (i) a linear 2-dimensional unicycle model from \cite{gracia2025efficient}, (ii) a nonlinear 3-dimensional unicycle  from \cite{gracia2024data}, in which the noise corresponds to both (nonlinear) \emph{coulomb} friction and additive noise on the yaw rate, and (iii) a multi-room temperature regulation benchmark from \cite{gracia2024temporal} with multiplicative noise and in a verification setting (fixed controller), where the number of rooms is $n \in \{2,3,4\}$.

To fairly compare the quality of the solutions yielded by all abstraction classes, we define the sets $\Gamma^\mi$ of all MI-MDP abstractions by considering both bounds on the probability of transitioning to each $s'\in S$, and also informed clusters (see Figure~\ref{fig:example3_b}).

Furthermore, while highly efficient implementations of RDP for IMDPs exist in C++ \cite{lahijanian2015formal} and Julia \cite{mathiesen2024intervalmdp}, no such implementations are available for the other three models: $2$I-MDPs, MI-MDPs, and SMDPs. To ensure a fair comparison, we implemented all algorithms and ran all benchmarks in MATLAB. We note that SMDPs could significantly benefit from a dedicated, optimized implementation. To perform RDP on MI-MDP abstractions, we used GUROBI. 
All experiments were conducted on a single thread of an Intel Core i7 3.6GHz CPU with 32GB of RAM.

Throughout all case studies, discrete set $S$ is a uniform partition of $X_\safe$, where each $s \in S_\safe$ region is an axis-aligned rectangle. Additionally, we let the partition $C$ of $W$ be as follows. When $W$ is bounded, we define $C$ by uniformly partitioning $W$ into axis-aligned regions $c\in C$. On the other hand, if $W$ is unbounded, we first define the ball $\widehat W := \{w\in W : \|w - w_0\|_\infty \le r_W\}$, for some center $w_0 \in \reals^d$ and radius $r_W >0$, which we uniformly partition, obtaining the axis-aligned rectangles $c_1,\dots,c_{|C|-1}$. Finally, we let $c_{|C|} := W\setminus \widehat W$.

We used the following metrics in our evaluations:
\begin{itemize}
    \item \emph{Tightness} ($e_\text{avg}$): the average of the difference between the probabilistic bounds $\underline p$ and $\overline p$ over all non-terminal states in $S_\mathrm{nt} = S_\safe \setminus S_\text{reach}$, i.e.,
    \begin{align}
        \label{eq: average error}
        e_\text{avg} := \frac{1}{|S_\mathrm{nt}|}\sum_{s\in S_\mathrm{nt}} (\overline p(s) - \underline p(s)).
    \end{align}

    \item \emph{Computation times}:
        \begin{itemize}
            \item[I.] $T_\text{abs}$: the total time taken to obtain the abstraction in minutes.
            \item[II.] $T_\text{syn}$: the total time taken to compute the probabilistic guarantees $\overline{p}$ and $\underline{p}$ as well as the optimal robust controller $\kappa$ in minutes.
        \end{itemize}

    \item \emph{Memory}: the total amount of memory used to store the abstraction in GB.

    \item \emph{Correctness} ($\overline{p}_\mathrm{avg}$, $\underline{p}_\mathrm{avg}$, and $P^\kappa_\text{avg}[\varphi_x]$): We first constructed the initial set $S_0 \subset S_\mathrm{nt}$ by randomly selecting $100$ states from $S_\mathrm{nt}$.  Then, we computed the following metrics: 
    \begin{itemize}
        \item[I.] $\overline{p}_\mathrm{avg}$ and $\underline{p}_\mathrm{avg}$: average, over $S_0$, theoretical guarantees of the lower and upper probabilistic bounds:
        \begin{align}
            \label{eq: average lower bound}
            \underline{p}_\text{avg} := \frac{1}{|S_0|}\sum_{s\in S_0} \underline p(s), \;\;
            \overline{p}_\text{avg} := \frac{1}{|S_0|}\sum_{s\in S_0} \overline p(s).
        \end{align}
        
        \item[II.] $P^\kappa_\text{avg}[\varphi_x]$: empirical reach-avoid probability obtained via Monte Carlo simulation of the closed-loop dynamics using $1000$ trajectories for each initial state $x_0\in s_0$, with $s_0\in S_0$.
    \end{itemize}    
\end{itemize}

\subsection{Benchmark Results}
\label{sec: overall results}


The detailed quantitative results of all case studies are provided in Table~\ref{tab:experiment_results}.
Here, we discuss the general trends, and then, in the subsequent subsections, we dive deeper into each case study. 






\begin{figure*}[t]
    \centering
    \begin{subfigure}[b]{0.22\linewidth}
        \centering
        \includegraphics[width=\linewidth]{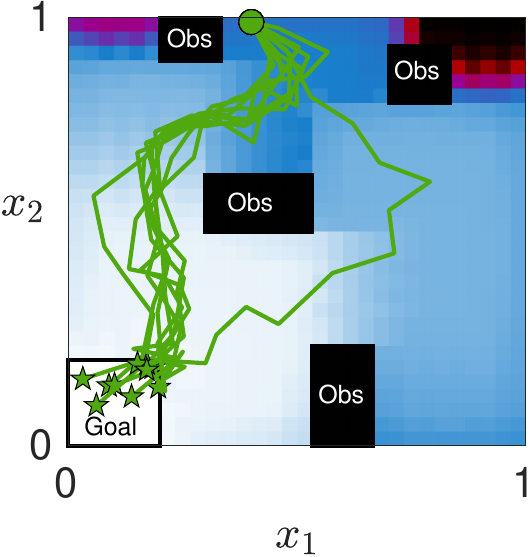}   
        \caption{IMDP}
        \label{fig:graphs_u2d_imdp}
    \end{subfigure}~
    \begin{subfigure}[b]{0.22\linewidth}
        \centering
        \includegraphics[width=\linewidth]{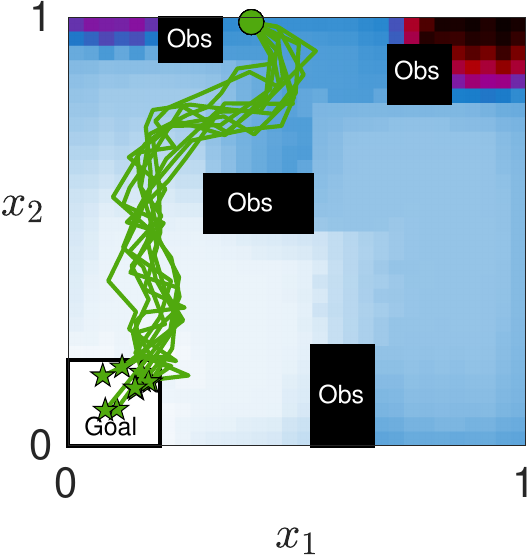}   
        \caption{2I-MDP}
        \label{fig:graphs_u2d_2layer}
    \end{subfigure}~
    \begin{subfigure}[b]{0.22\linewidth}
        \centering
        \includegraphics[width=\textwidth]{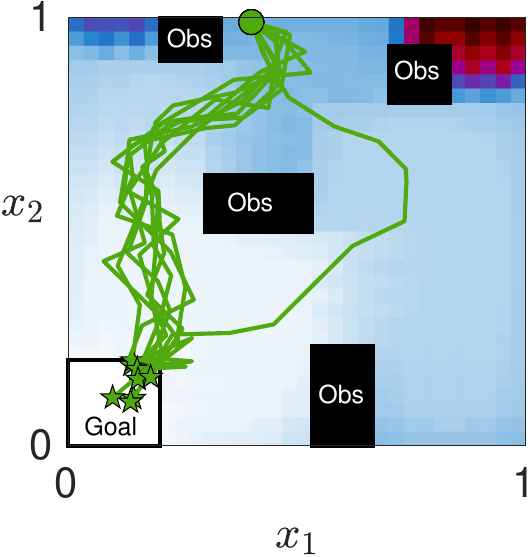}   
        \caption{MI-MDP}
        \label{fig:graphs_u2d_mimdp}
    \end{subfigure}~
    \begin{subfigure}[b]{0.2725\linewidth}
        \centering
        \includegraphics[width=\linewidth]{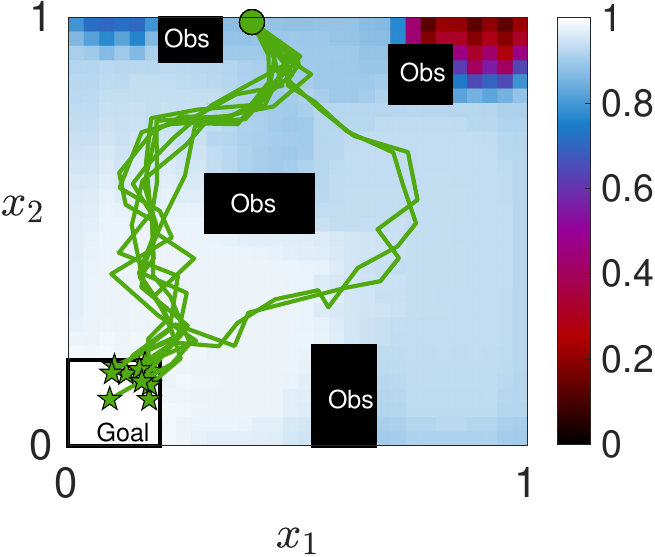}
        \caption{SMDP}
        \label{fig:graphs_u2d_smdp}
    \end{subfigure}
    \caption{
    $2$D-unicycle benchmark:  background color indicate probabilistic guarantee $\underline p(x)$ from each initial state, and the green lines are sample trajectories of the closed-loop system from the same initial state. The results correspond to rows $1$-$4$ in Table~\ref{tab:experiment_results}.
    }
    \label{fig:plots_u2d}
\end{figure*}
\begin{figure*}[t]
    \centering
    %
    \begin{subfigure}[b]{0.24\linewidth}
        \centering
        \includegraphics[width=\linewidth]{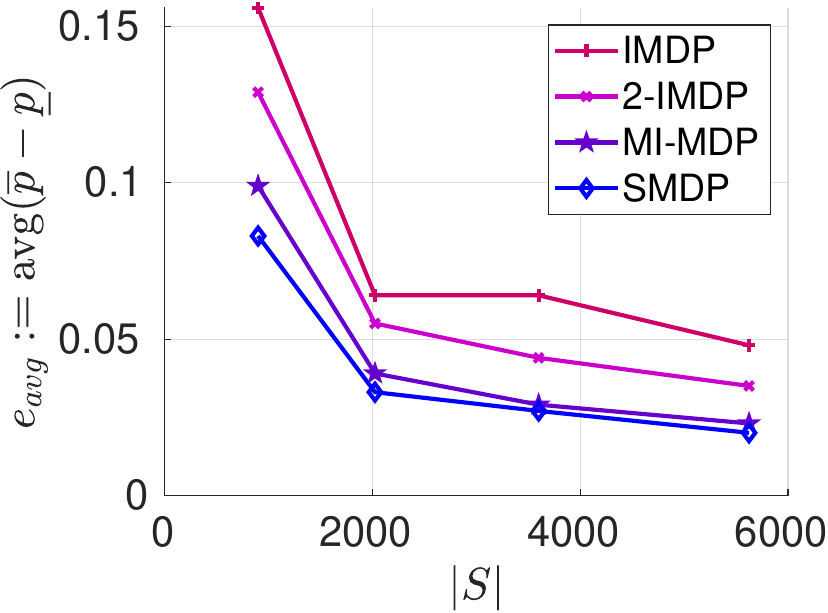}    
        \caption{Tightness}
        \label{fig:graphs_a}
    \end{subfigure}~
    \begin{subfigure}[b]{0.23\linewidth}
        \centering
        \includegraphics[width=\linewidth]{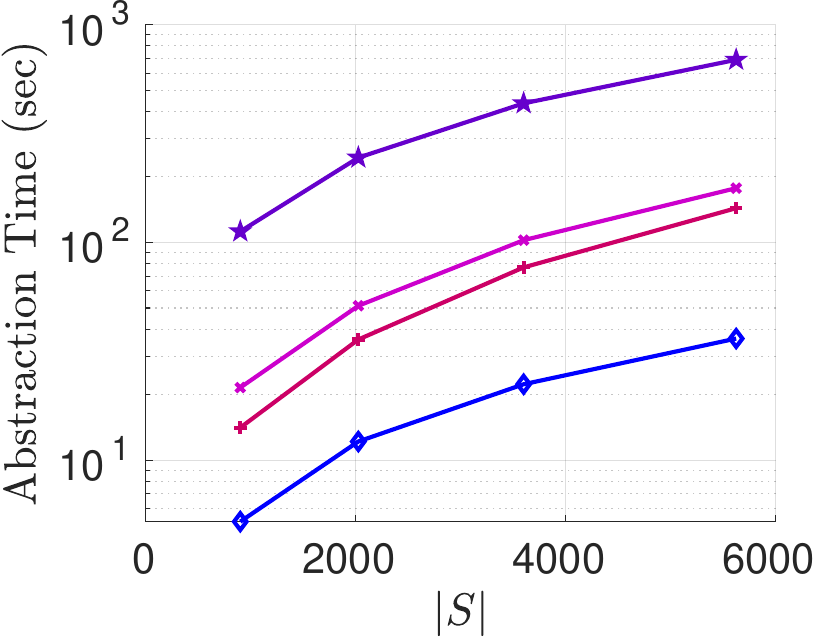}
        \caption{Abstraction time}
        \label{fig:graphs_b}
    \end{subfigure}~
    \begin{subfigure}[b]{0.24\linewidth}
        \centering
        \includegraphics[width=\textwidth]{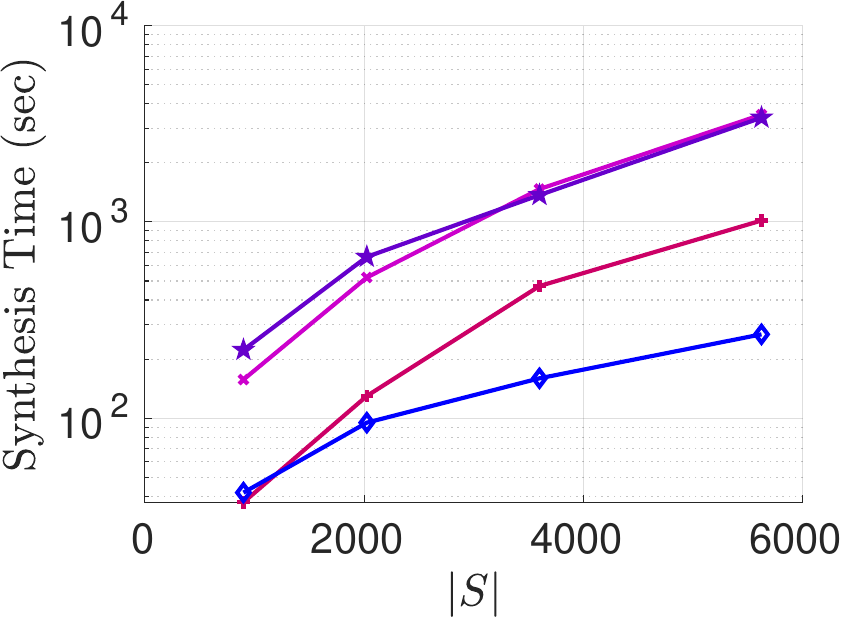}
        \caption{Synthesis time}
        \label{fig:graphs_c}
    \end{subfigure}~
    \begin{subfigure}[b]{0.24\linewidth}
        \centering
        \includegraphics[width=\textwidth]{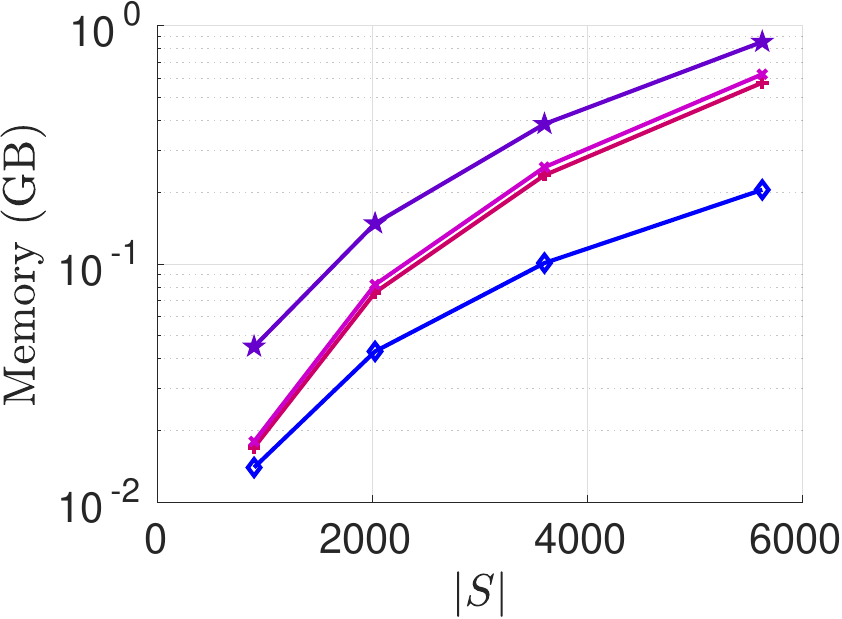}
        \caption{Memory}
        \label{fig:graphs_d}
    \end{subfigure}
    \caption{$2$D-unicycle benchmark: effect of the granularity of the partition $S$ (rows $1$-$16$ in Table~\ref{tab:experiment_results}).}
    \label{fig:graphs_u2d}
\end{figure*}

\textit{Correctness: }
Across all experiments, the empirical results align with the theoretical guarantees: we observe that $P^\kappa_\text{avg}[\varphi_x] \in [\underline{p}_\text{avg}, \overline{p}_\text{avg}]$, confirming the correctness of all approaches. 
Moreover, since $P^\kappa_\text{avg}[\varphi_x]$ consistently lies closer to $\overline{p}_\text{avg}$ than to $\underline{p}_\text{avg}$, we conclude that the abstraction-based approaches are more conservative in their lower-bound estimates.

\textit{Tightness}:
In addition to correctness, achieving tight satisfaction bounds is a key goal in formal synthesis. Our results show that MI-MDPs and especially SMDPs provide significantly tighter guarantees, as reflected in the lower values of $e_\text{avg}$.

\textit{Computation Time}:
SMDPs consistently demonstrate fast abstraction construction across all case studies. They achieve the smallest abstraction times compared to MI-MDPs, 2I-MDPs, and IMDPs. While control synthesis for SMDPs can be slightly slower than for IMDPs when the state partition is coarse (e.g., $|S| = 901$), SMDPs generally match or outperform other models as the partition becomes finer. Overall, SMDPs provide a favorable trade-off between computation time and abstraction tightness.

\textit{Memory}:
Among all models, IMDPs and SMDPs require the least memory. In higher-dimensional settings, IMDPs use the smallest amount of memory overall. However, this memory efficiency comes at the cost of reduced tightness, especially when compared to SMDPs.

Overall, SMDP abstractions offer the best trade-off between tightness, computation time, and memory usage across all case studies.

\subsection{$2$D Unicycle}

The system dynamics are given in \cite{gracia2025efficient}, but here we consider a noisier setting in which the covariance of the Gaussian noise is $\mathrm{diag}(0.3^2, 0.3^2)$
, and the time discretization $\Delta t = 0.1$
. We let the sets $X_\text{reach}, X_\text{safe} \subset [0,1]^2$ be as shown in Figure~\ref{fig:plots_u2d}. Since the disturbance is unbounded, we obtain the partition $C$ as explained before with $w_0 = 0$ and $r_W = 2.1$.


Figure~\ref{fig:plots_u2d} illustrates the reach-avoid probabilistic guarantees $\underline p$, indicated by the background color, for each initial state across the different abstraction classes. The figures also include $10$ Monte Carlo simulations of the closed-loop system from a selected initial state. Observe that, while all abstraction-based methods demonstrate strong empirical performance, SMDPs consistently yield the highest values of $\underline p$, followed by MI-MDPs, $2$I-MDPs, and IMDPs.

\begin{figure*}[t]
    \centering
    \begin{subfigure}[b]{0.23\linewidth}
        \centering
        \includegraphics[width=\linewidth]{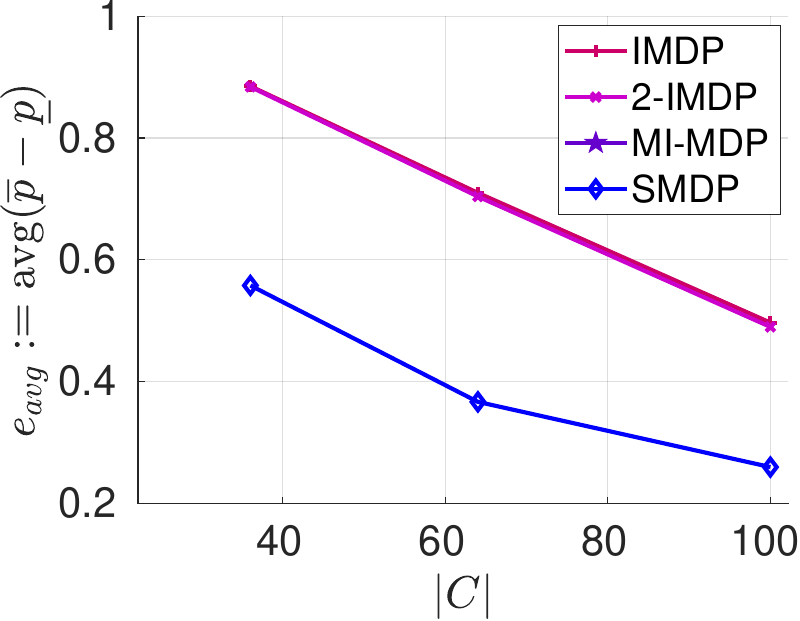}        
        \caption{Tightness}
        \label{fig:graphs_u3d_a}
    \end{subfigure}~
    \begin{subfigure}[b]{0.248\linewidth}
        \centering
        \includegraphics[width=\linewidth]{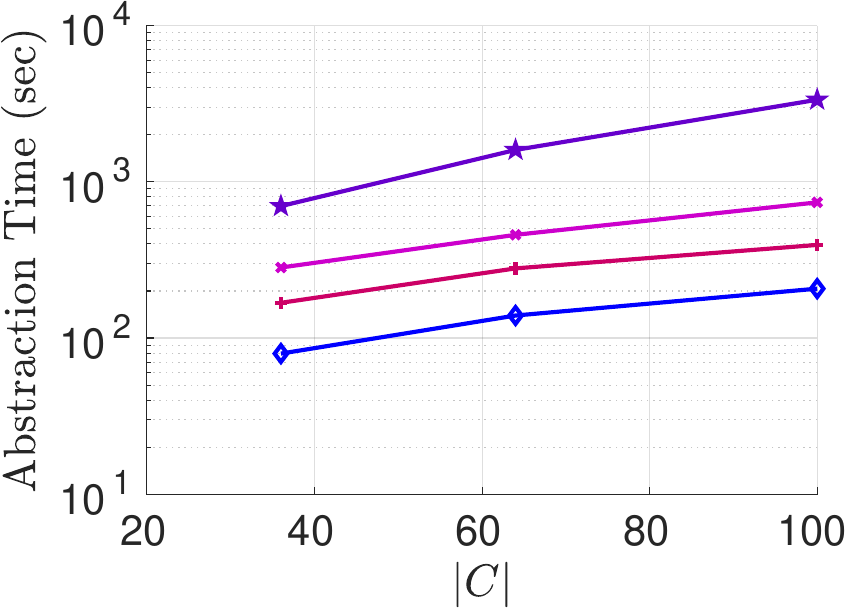}
        \caption{Abstraction time
        }
        \label{fig:graphs_u3d_b}
    \end{subfigure}~
    \begin{subfigure}[b]{0.24\linewidth}
        \centering
        \includegraphics[width=\textwidth]{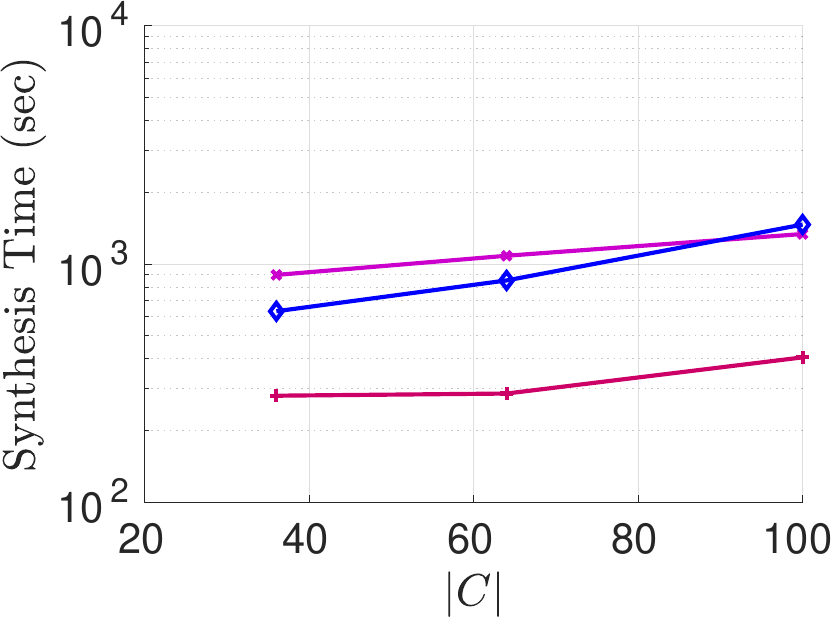}
        \caption{Synthesis time
        }
        \label{fig:graphs_u3d_c}
    \end{subfigure}~
    \begin{subfigure}[b]{0.248\linewidth}
        \centering
        \includegraphics[width=\textwidth]{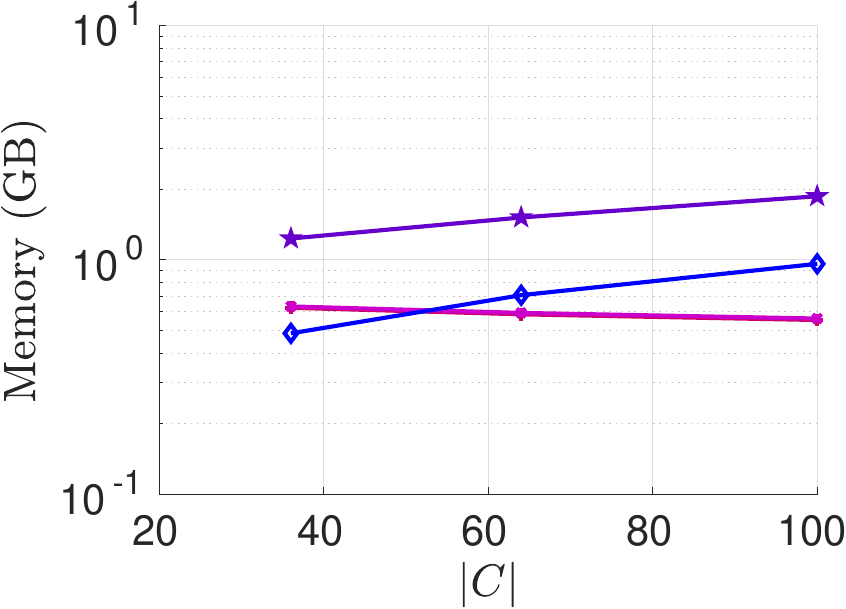}
        \caption{Memory
        }
        \label{fig:graphs_u3d_d}
    \end{subfigure}
    \caption{$3$D-unicycle benchmark: effect of the granularity of the partition $C$ (rows
    $17$-$28$ in Table~\ref{tab:experiment_results}). The $e_\text{avg}$ and synthesis time for MI-MDP in \ref{fig:graphs_u3d_a} and \ref{fig:graphs_u3d_c} are unavailable because control synthesis timed out.}
    \label{fig:graphs_u3d}
\end{figure*}
\begin{figure*}[t]
    \centering
    \begin{subfigure}[b]{0.3\linewidth}
        \centering
        \includegraphics[width=\linewidth]{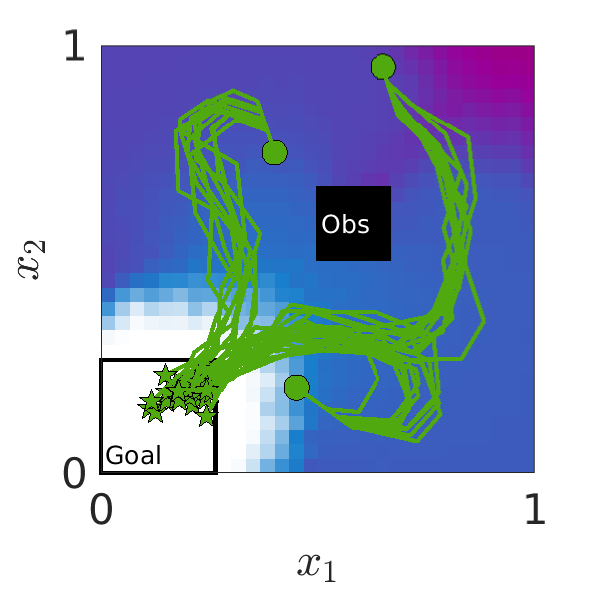}   
        \caption{$|C| = 37$}
        \label{fig:graphs_u3d_smdp_6}
    \end{subfigure}~
    \begin{subfigure}[b]{0.3\linewidth}
        \centering
        \includegraphics[width=\linewidth]{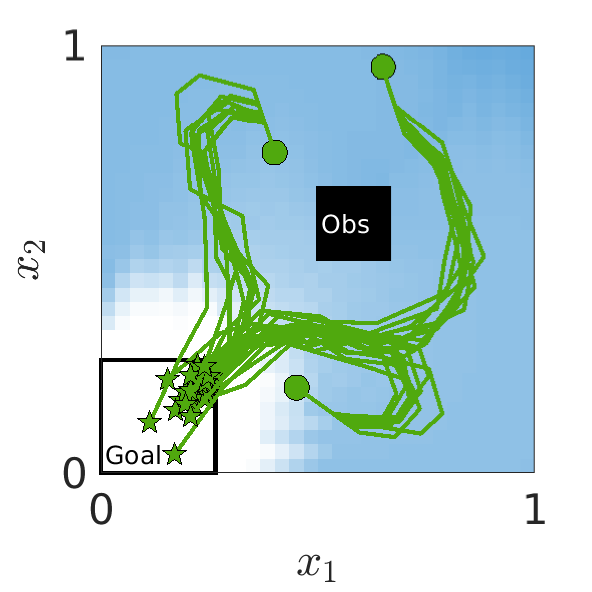}   
        \caption{$|C| = 65$}
        \label{fig:graphs_u3d_smdp_8}
    \end{subfigure}~
    \begin{subfigure}[b]{0.35\linewidth}
        \centering
        \includegraphics[width=\textwidth]{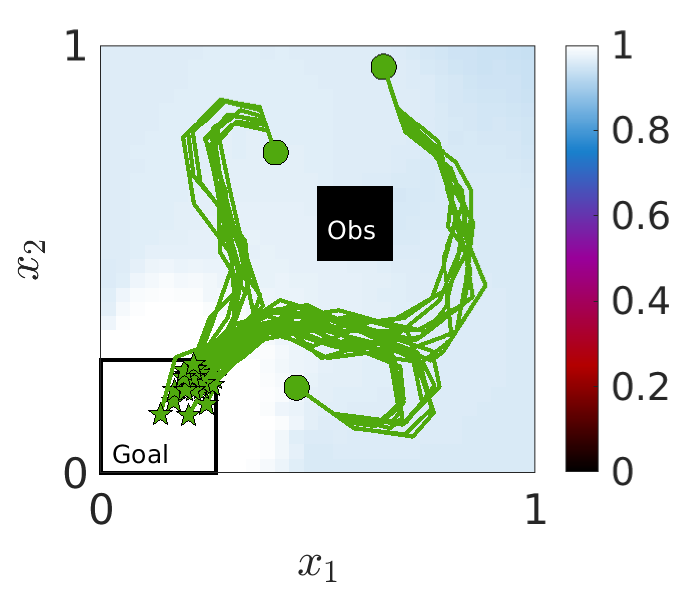}   
        \caption{$|C| = 101$}
        \label{fig:graphs_u3d_smdp_10}
    \end{subfigure}~
    \caption{
    3D-unicycle benchmark results: 
    effect of the granularity of the disturbance partition $C$ on the guarantees provided by the SMDP abstraction (rows $20, 24$ and $28$ in Table~\ref{tab:experiment_results}).
    Background color indicates probabilistic guarantee $\underline p(x)$ from each initial state, and the green lines are sample trajectories of the closed-loop system from the same three initial states.
    }
\label{fig:plots_u3d}
\end{figure*}
%



Figure~\ref{fig:graphs_u2d} shows the impact of the discretization size $S$ on four key metrics: tightness, computation time (for both abstraction and control synthesis), and memory usage, across all abstraction classes. Note that Figures~\ref{fig:graphs_b}-\ref{fig:graphs_c} are in logscale. As expected, refining the partition $S$ leads to smaller average error 
$e_\text{avg}$, resulting in tighter abstractions and more precise guarantees. However, this refinement comes at the cost of increased abstraction complexity and longer computation times.

Consistent with Theorem~\ref{thm:tightness_guarantees}, SMDPs yield the tightest results, followed by MI-MDPs, 2I-MDPs, and finally IMDPs. Notably, SMDPs outperform MI-MDPs even when both use the same (informed) clusters. In terms of computation time, SMDPs are the most efficient when the partition is sufficiently fine. For example, at 
$|S|=901$, IMDPs are slightly faster in control synthesis, but SMDPs become more efficient as the partition is further refined.

SMDPs also require the least memory in this case study. As discussed in Section~\ref{sec:memory}, this is largely due to the use of a fine partition $S$ and the low dimensionality of the system.


\subsection{$3$D Unicycle}

The system dynamics are given in \cite{gracia2024temporal}, although here we consider that the Gaussian disturbance has a remarkably higher covariance of $\mathrm{diag}(0.5^2, 0.5^2)$
. We let the sets $X_\text{reach}, X_\text{safe} \subset [0,1]^2\times [0, 2\pi]$ be as shown in Figure~\ref{fig:plots_u3d}. Since the disturbance is unbounded, we obtain the partition $C$ as explained before with $w_0 = [0.4, 0]^T$ and $r_W = 2$.

Figure~\ref{fig:graphs_u3d} shows the evaluation metrics for all abstraction classes as a function of $|C|$. A general trend is that refining $C$ reduces $e_\text{avg}$, thereby improving the tightness of the abstraction. Consistent with the 2D unicycle case study, SMDPs achieve the lowest $e_\text{avg}$, followed by MI-MDPs, 2I-MDPs, and finally IMDPs. Notably, while MI-MDPs incur the highest computation times, SMDPs yield more accurate results with the shortest abstraction times and synthesis times comparable to those of 2I-MDPs—and relatively close to IMDPs.

\begin{figure*}[t]
    \centering
    %
    \begin{subfigure}[b]{0.23\linewidth}
        \centering
        \includegraphics[width=\linewidth]{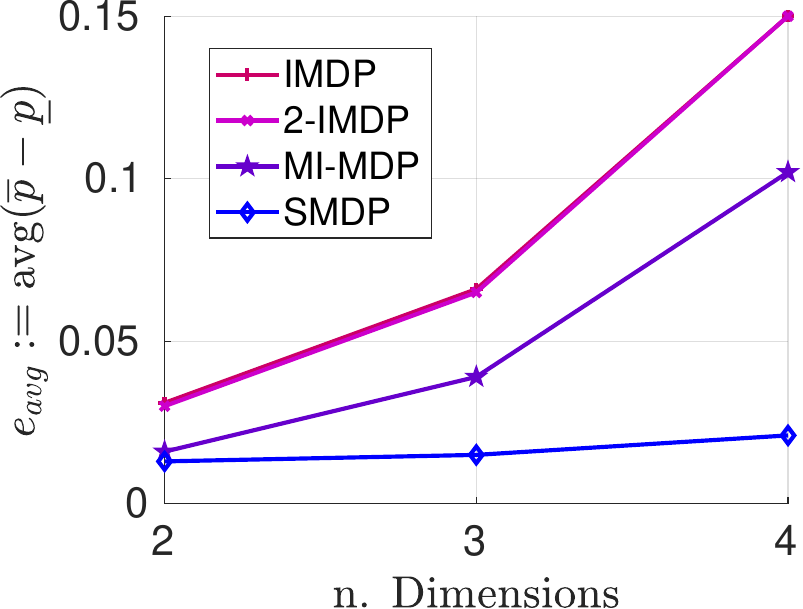}    
        \caption{Tightness}
        \label{fig:graphs_temp_a}
    \end{subfigure}~
    \begin{subfigure}[b]{0.225\linewidth}
        \centering
        \includegraphics[width=\linewidth]{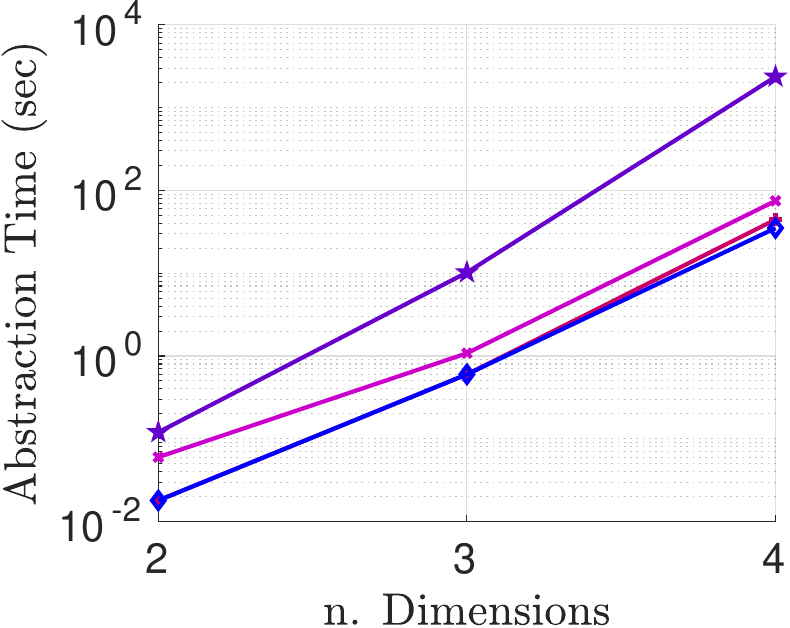}
        \caption{Abstraction time}
        \label{fig:graphs_temp_b}
    \end{subfigure}~
    \begin{subfigure}[b]{0.24\linewidth}
        \centering
        \includegraphics[width=\textwidth]{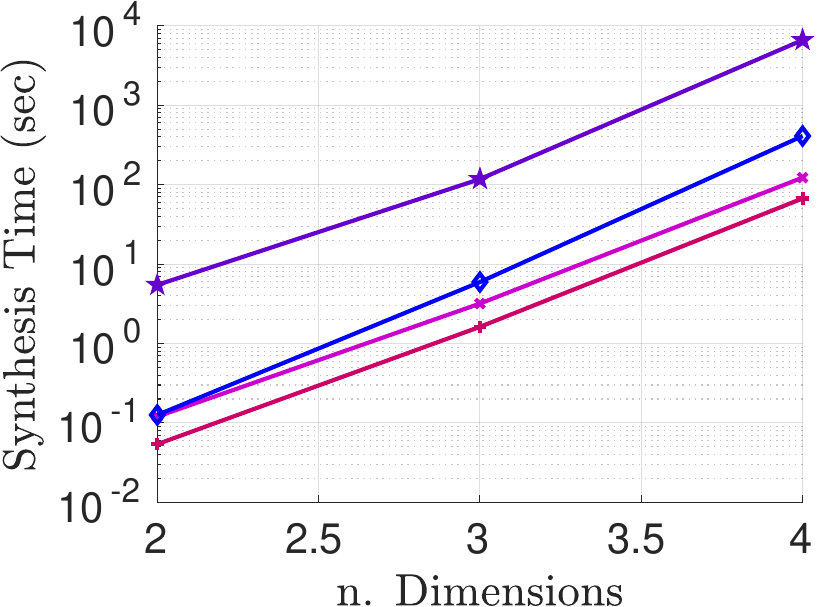}
        \caption{Synthesis time}
        \label{fig:graphs_temp_c}
    \end{subfigure}~
    \begin{subfigure}[b]{0.24\linewidth}
        \centering
        \includegraphics[width=\textwidth]{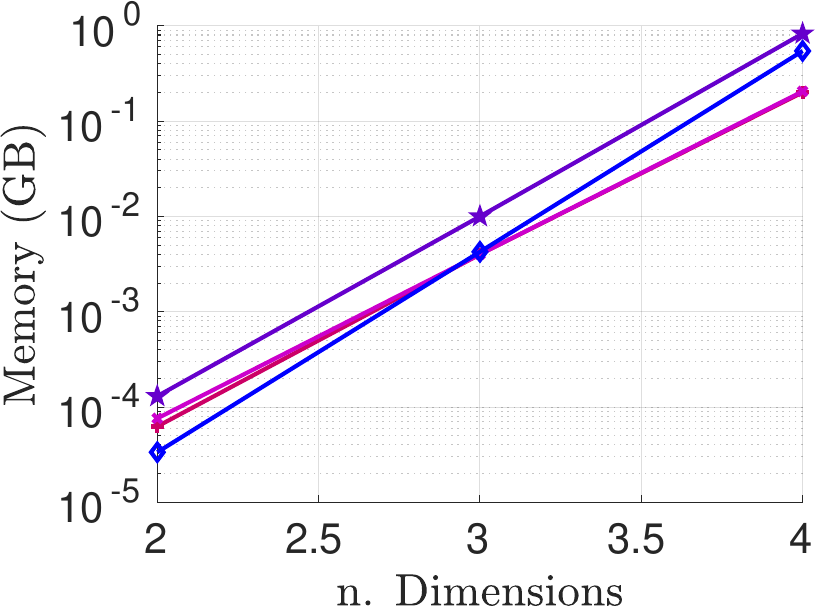}
        \caption{Memory}
        \label{fig:graphs_temp_d}
    \end{subfigure}
    \caption{
    Room temperature benchmark results:
    effect of increasing the dimension $n$ 
    (rows $29$-$40$ in Table~\ref{tab:experiment_results}).}
    \label{fig:graphs_temp}
\end{figure*}

Although a finer discretization $C$ increases the memory complexity of SMDPs and MI-MDPs due to a larger number of clusters, this is not the case for IMDPs and 2I-MDPs, where refining $C$ actually reduces memory usage. This is because the cluster structure in IMDPs and 2I-MDPs remains fixed, while a finer $C$ results in tighter overapproximations of the reachable sets $\mathrm{Reach}(s,a,c)$, thereby reducing the set of possible successor states. Interestingly, for $|C| = 37$, SMDPs require less memory than both IMDPs and 2I-MDPs, as explained in Section~\ref{sec:memory}. In contrast, MI-MDPs, due to the need to store a significantly larger number of transition probability intervals, report the highest memory consumption.

Figure~\ref{fig:plots_u3d} further illustrates the reach-avoid probabilistic guarantee $\underline{p}$ obtained using the SMDP abstraction for different values of $|C|$. It also shows Monte Carlo simulations of trajectories from three selected initial states. All simulated trajectories satisfy the reach-avoid specification, and increasing the resolution of $C$ leads to higher satisfaction probabilities.

\subsection{Multi-Room Temperature Regulation}

The system dynamics are given in \cite{gracia2024temporal}, although here we consider a small control authority by letting $b_u$ be multiplied by a factor of $0.8$
. For this case study, instead of synthesizing a controller to enforce a reach-avoid specification, we consider the problem of verifying safety of a given controller, i.e., the system remains in $X_\safe$ on a given time horizon. We consider a controller given as a look-up table, and a time horizon of $15$ time steps. Since the disturbance is Gaussian and unbounded, we obtain the partition $C$ as explained above with $w_0 = 0$ and $r_W = 0.0295$.



Figure~\ref{fig:graphs_temp} shows how increasing the dimension $n$ of the state space, corresponding to the number of rooms in this case study, affects tightness, computation time, and memory usage across all abstraction classes. As in the previous case studies, we observe that SMDPs consistently yield the tightest results, i.e., the smallest $e_\text{avg}$, while also achieving the shortest abstraction times. However, their synthesis times are slightly higher than those of IMDPs and 2I-MDPs.

We also observe that when $n$ is small, SMDPs exhibit the lowest memory usage among all models. As $n$ increases, however, their memory usage grows more rapidly, eventually surpassing all models except MI-MDPs at higher dimensions. This trend is consistent with the results from the previous two case studies and supports the discussion in Section~\ref{sec:memory}. Overall, SMDPs effectively limit the growth of $e_\text{avg}$ with increasing $n$, albeit at the cost of a faster increase in memory complexity compared to the other abstraction methods.



\section{Conclusion and Future Work}


In this work, 
we introduced two abstraction-based approaches, namely MI-MDPs and SMDPs, for controller synthesis in nonlinear stochastic systems, both aimed at reducing the conservatism of existing methods.
MI-MDPs generalize prior abstractions such as IMDPs by allowing multiple, overlapping clusters, leading to tighter guarantees at the cost of increased memory and computation. In contrast, SMDPs are shown to be at least as tight as any MI-MDP under the same discretization, while offering significantly lower abstraction times and synthesis costs comparable to IMDPs.
Our extensive empirical evaluation supports the theoretical findings and further demonstrates that SMDPs effectively mitigate the growth in conservatism with increasing system dimensionality.

Our ongoing research focuses on efficient data structures to reduce the computational complexity of SMDP abstractions. Another direction is the extension of SMDP abstractions to data-driven setting, where both vector field $f$ and disturbance distribution $P_W$ are unknown. 
While Theorem~\ref{thm:multi_imdp_vs_mdp_st} and the case studies in Section~\ref{sec:case_studies} show that SMDPs yield tighter guarantees than MI-MDPs in model-based scenarios, MI-MDPs may offer advantages when the abstraction must be constructed from sampled system trajectories.

\begin{ack}
We thank the anonymous reviewer of our L4DC paper \cite{gracia2024temporal} for mentioning the work of \cite{yu2025planning} on SMDPs.
\end{ack}

\bibliographystyle{plain}        
\bibliography{Ref}           

\end{document}